\documentclass[pdftex,twocolumn,epjc3]{svjour3}          

\RequirePackage[T1]{fontenc}

\smartqed  
\RequirePackage{caption}
\RequirePackage{graphicx}
\RequirePackage{mathptmx}      
\RequirePackage{flushend}
\RequirePackage[numbers,sort&compress]{natbib}
\RequirePackage[colorlinks,citecolor=blue,urlcolor=blue,linkcolor=blue]{hyperref}

\journalname{Eur. Phys. J. C}

\begin{document}

\title{The nonleptonic decays of $b$-flavored mesons to $S$-wave charmonium and charm meson states}


\author{Kalpalata Dash\thanksref{e1,addr1} \and
        Lopamudra Nayak\thanksref{e2,addr1} \and
        P. C. Dash\thanksref{addr1}\and
        Rabinarayan Panda\thanksref{addr1}\and 
        Susmita Kar\thanksref{addr2}\and
        N. Barik\thanksref{addr3} 
}

\thankstext{e1}{e-mail: kalpalatadash982@gmail.com}
\thankstext{e2}{e-mail: lopalmn95@gmail.com}

\institute{Department of Physics, Siksha $'O'$ Anusandhan Deemed to be University, Bhubaneswar-751030, India\label{addr1}
          \and
          Department of Physics, Maharaja Sriram Chandra Bhanja Deo University, Baripada-757003, India\label{addr2}
          \and
          Department of Physics, Utkal University, Bhubaneswar-751004, India\label{addr3}
}
\maketitle
\begin{abstract}
The detection of radially excited heavy meson \\states in recent years and measurement of heavy meson decays, particularly $B_c^+\to J/\psi D_s^+$ and $B_c^+\to J/\psi D_s^{*+}$, by the LHCb and ATLAS Collaborations, have aroused a lot of theoretical interest in the nonleptonic decays of $b$-flavored mesons. In this paper, we study the exclusive two-body nonleptonic $\bar{B}^0$, $\bar{B_s^0}$, $B^-$ and $B_c^-$-meson decays to two vector meson ($V_1(nS)V_2$) states. Assuming the factorization hypothesis, we calculate the weak-decay form factors from the overlapping integrals of meson wave functions, in the framework of the relativistic independent quark (RIQ) model. We find a few dominant decay modes: $B^-\to D^{*0}\rho^-$, $\bar{B^0}\to D^{*+}\rho^-$, $\bar{B_s^0}\to D_s^{*+}\rho^-$, $B^-\to J/\psi K^{*-}$ and $B_c^-\to J/\psi D_s^{*-}$ with predicted branching fractions of 1.54, 1.42, 1.17, 0.53 and 0.52 (in $\%$), which are experimentally accessible. The predicted branching fractions for corresponding decay modes to excited ($2S$) states, obtained in the order ${\cal O }(10^{-3}-10^{-4})$ lie within the detection accuracy of the current experiments at LHCb and Tevatron. The sizeable $CP$-odd fractions predicted for $B_c^-$-meson decay to two charmful states: $D^{*0}D^{*-}_{(s)}$ and $\bar{D}^{*0}D^{*-}_{(s)}$ indicate significant $CP$-violation hinting at the so-called new physics beyond the standard model.	
\end{abstract}
\section{Introduction}
The experimental probes over the last two decades in the $b$-flavored heavy meson ($B, B_s, B_c$) sector have led to the discovery of many excited states which include the radially excited charmonium $\psi(2S)$ and $\eta_c(2S)$ states by the Belle Collaborations \cite{A1}. In the heavy-light meson sector, a number of charm meson states such as $D_{s1}^*(2710)^\pm$ \cite{A2} by Belle, $D(2550)^0$ \cite{A3}, and $D_J(2580)^0$ \cite{A4} by the LHCb, $D^*(2640)^\pm$ \cite{A5} by Delphi and $D^*(2650)^0$ \cite{A4} and $D_1^*(2680)^0$ \cite{A6} by LHCb have also been discovered, which can be identified as $D_s^*(2S)^\pm$ \cite{A7}, $D(2S)^0$ \cite{A8}, $D^*(2S)^\pm$ \cite{A8} and $D^*(2S)^0$ state, respectively. Recently LHCb discovered $D_1^*(2600)$ \cite{A3} which might be the same object as $D^*(2650)^0$ and $D_1^*(2650)$. The current RUN-II at Tevatron, RUN-III at CERN LHC, and the $e^+e^-$ collider activities at Belle-II are all designed to boost the measurement scenario in heavy flavor physics. Specially designed detectors at BTeV and LHCb, dedicated to enhancing the event accumulation rates, are expected to yield high statistics $b$-flavored ($B, B_s$) events and the $B_c$-events, in particular, at the rate $\sim 10^{10}$ events per annum; providing a fascinating area of research in $B$-physics. The measurement of $B, B_s$-meson decays by BaBar, Belle, and LHCb Collaborations \cite{A9,A10,A11} and recent measurement of $B_c$-meson decays: $B_c\to J/\psi D_s$ and $B_c\to J/\psi D_s^*$, performed by LHCb \cite{A12} and ATLAS \cite{A13} Collaborations have aroused a great deal of theoretical interest in nonleptonic decays of heavy-flavored mesons. The study of nonleptonic decays of heavy mesons is important as it helps in probing the interplay of QCD and electroweak interactions, determining the Cabibbo-Kobayashi-Maskawa (CKM) matrix elements, testing predictions of the standard model (SM), and exploring new physics beyond SM. 
 \par The nonleptonic decays of $b$-flavored mesons have been widely studied in different theoretical and phenomenological model approaches (see the classified bibliography of Ref.14). Most of these earlier studies refer to the $b$-flavored $(B, B_s, B_c)$-meson decays to ground states of charmonium and charm mesons. A number of theoretical attempts have also been made in this sector, yielding predictions on decay modes to radially excited states. A few noteworthy among them include studies based on the non-relativistic potential model (NRPM) using the Bathe-Salpeter equation \cite{A15,A16}, relativistic constituent quark model (RCQM) based on the Bathe-Salpeter formalism \cite{A17}, improved Bathe-Salpeter (IB\\S) approach \cite{A18,A19}, relativistic quark model (RQM) \cite{A20,A21,A22,A23}, ISGW2 quark model \cite{A24}, covariant confined quark model (CCQM) \cite{A25}, relativistic quark model (RQM) using quasi-potential approach \cite{A26}, QCD relativistic potential model (QCDRPM)\cite{A27}, QCD factorization \cite{A28}, perturbative QCD (pQCD) \cite{A29}, covariant light-front quark model (CLFQM) \cite{A30}, light-cone sum rule (LCSR) \cite{A31} and the heavy quark sum rules (HQS) \cite{A32} etc. The recent predictions of branching fraction for $B_c\to J/\psi D_s^*(2S) \simeq 1.75\times10^{-2}$, $B_c\to \psi(2S)\\ D^*_s \simeq 2.72\times10^{-3}$ based on improved Bathe-Salpeter approach \cite{A18,A19}, $B_s\to D^*_s(2S) D^*_s\simeq1.7\times10^{-3}$ in RQM \cite{A26}, $B_c^-\to\psi(2S)\rho^-\simeq 1.1\times10^{-3}$, $B_c\to\psi(2S)D^*_s\simeq1.2\times10^{-3}$ in ISGW2 quark model \cite{A24}, $B_c^+\to D^{*+}\bar{D}^{*0}\simeq5.14\times10^{-3}$ in  QCDRPM \cite{A27} etc. are accessible in the LHCb experiment. The branching fractions for decay modes: $B_c^-\to\psi(2S)\\\rho^-$ \cite{A16}, $B_s\to D_s^*\rho^-$ and $B_c^-\to \psi(3S)\rho^-$\cite{A18}, etc. predicted in the range $\sim10^{-4}$, lie within the detection accuracy of the current experiments and can be measured in near future.
\par The analysis of nonleptonic decays is notoriously non-trivial as it is strongly influenced by confining color forces and it involves matrix elements of local four-quark operators in the non-perturbative QCD approach, the mechanism of which is not yet clear in the SM framework. Ignoring the weak annihilation contribution, the transition amplitudes can be conveniently described in the so-called naive factorization approximation \cite{A17,A20,A21,A22,A23,A27,A33,A34,A35,A36,A37,A38,A39,A40,A41,A42,A43,A44,A45,A46,A47,A48,A49,A50,A51,A52,A53,A54,A55}, which works reasonably well in the nonleptonic $b$-flavored meson decays, where the quark-gluon sea is suppressed in the heavy quarkonium \cite{A47,A48,A49,A50}. In this approach, the transition matrix element of local four-quark operators is factorized into two single current matrix elements. One of the factorized amplitudes, in which the decaying parent meson is connected to one final meson state, can be covariantly expanded in terms of Lorentz invariant weak form factors as in the case of semileptonic decays. The other factorized amplitude, where the vacuum is connected to the second final meson state, can be parametrized in terms of meson decay constants that describe the leptonic decays. The description of the nonleptonic decay process is thus reduced to the calculation of weak decay form factors in the framework of a suitable phenomenological model.
 \par Bjorken's intuitive argument on color transparency in his pioneering work \cite{A51}, theoretical development based on the QCD approach in the $\frac{1}{N_c}$ limit \cite{A52} and the heavy quark effective theory (HQET) \cite{A53}, etc. justify the naive factorization approximation, where strong-interaction effects such as the final-state interaction, rescattering of the final state hadrons and the renormalization-point dependence of amplitudes are shown to be marginal \cite{A56}. Discovery of excited charmonium and charm meson states and prediction of $b$-flavored meson decays to the ground and radially excited states by different theoretical approaches, inspired our group to predict energetic nonleptonic $B$ and $B_c\to PP, PV, VP$ decays \cite{A33,A34} as well as their decays to two vector meson ($VV$) ground states \cite{A35,A36}, within the framework of our relativistic independent quark (RIQ) model. Here $P$($V$) refers to a pseudoscalar (vector) meson state. The nonleptonic $B_c$-decays to an $S$-wave charmonium and a light or charm meson state \cite{A55} have also been predicted by our group in good comparison with the experiment and other SM predictions. \par Note here that the approach based on naive factorization approximation may be justified in the analysis of energetic nonleptonic decays of $B_F\to PP, PV$ type, with the quark flavor $F\to d, u, s, c$; where the strong interaction effects such as the final state interaction, rescattering of final state meson as well as the renormalization point dependence of factorized amplitude have been shown to be marginal  \cite{A57}. Such an approach, however, may not hold up well in the description of $B_F\to V_1(nS)V_2$ decays, where both the final state mesons being heavy, are expected to be in the  region close to zero recoil point. Here also both the longitudinal and transverse polarization components contribute to decay amplitude  which can be measured experimentally. From the naive counting rules, the longitudinal polarization fraction in this sector is expected to dominate over transverse components which can be checked as well. The nonleptonic $B_F$-decays to two charmful vector meson states are of special interest as they provide valuable information which is different from cases with light meson productions. For example, the evaluation of $CP$-asymmetries in $B^-_c$ decays: $B^-_c\to \bar{D}^{*0}D_{(s)}^{*-}$ and $B^-_c\to D^{*0}D_{(s)}^{*-}$ provides an important clue in testing the SM predictions and exploring new physics beyond SM.
 \par In this paper we would like to extend the applicability of our RIQ model to study, within factorization approximation, the nonleptonic $B_F\to V_1(nS)V_2$ decays to $S$- wave charmonium and charm vector meson states $(nS)$ along with a light or a heavy-light meson state, where $n=1, 2, 3$. We ignore the decay channels involving higher $(4S)$ charmonium and charm meson states since their properties are still not understood well. We adopt here the general formalism used in Ref. \cite{A55}. In the present study we consider the contribution of the current-current operators \cite{A58} only in calculating the tree-level diagram, expected to be dominant in $B_F\to V_1(nS)V_2$ decays. The contribution of the penguin diagram may be significant in the evaluation of $CP$-violation and search for new physics beyond SM, but its contribution to these decay amplitudes is considered less significant. In fact, the QCD and electroweak penguin operators' contribution have been shown \cite{A59,A60,A61,A62} negligible compared to the contribution of current-current operators in these decays due to serious suppression of CKM matrix elements. The Wilson's coefficients of penguin operators being very small, their contribution to decay amplitude is only relevant in rare decays, where the tree-level contribution is either strongly CKM-suppressed as in $\bar{B}\to \bar{K}^*\pi$ or matrix elements of current-current operators do not contribute at all as in the case of rare decays: $\bar{B}\to \bar{K}^*\gamma$ and $\bar{B^0}\to \bar{K}^0\phi$ \cite{A58}.
 \par The rest of the paper is organized as follows. In the following section, we present a general remark on the factorization approximation and discussed the factorized amplitudes of the nonleptonic decay. In Section 3, we obtain the model expressions for invariant weak-decay form factors and the factorized transition amplitudes. Section 4 is devoted to the numerical results and discussion and Section 5 encompasses our brief summary and conclusion. A brief review of the model conventions, wave-packet representation of the meson state and momentum probability amplitudes of the constituent quarks inside the meson bound-state are given in the Appendix.  
\section{Factorization approximation and nonleptonic transition amplitude}   
 The transition amplitude for two-body nonleptonic transition:
  $B_F\to V_1(nS)V_2$ can be written as 
  \begin{equation}
 \resizebox{.9\hsize}{!}{$A(B_F\to V_1(nS)V_2)=\langle V_1(nS)V_2|{\cal H}|B_F \rangle ={\frac{ G_F}{\sqrt{2}}}\sum_{i}\lambda_i C_i(\mu){\langle{\cal O}\rangle}_i,$}   
\end{equation}
  where $G_F$ is the Fermi Coupling constant, $\lambda_i$ the CKM factor, $C_i$ is the Wilson coefficients and ${\langle{\cal O}\rangle}_i$ is the matrix element of local four-quark operators. In the factorization approximation, the matrix element of the local four-quark operator is factorized into two single-particle matrix elements of quark current as
  \begin{equation}
 \resizebox{.9\hsize}{!}{${\langle V_1(nS)V_2|{\cal H}|B_F \rangle}_i={\langle V_2|J^\mu|0\rangle}{\langle V_1(nS)|J_\mu|B_F\rangle}+{(V_1(nS)\leftrightarrow V_2)},$} 
\end{equation}
where $J_\mu \equiv V_\mu-A_\mu$  is the vector-axial vector current.\\
 \par The difficulty inherent in such an approach is that Wilson's coefficients $C_i(\mu)$, which include the short distance QCD effect between $\mu=m_N$ and $\mu=m_b$ are $\mu$ scale and renormalization scheme dependent, whereas ${\langle{\cal O}\rangle}_i$ are $\mu$ scale and renormalization scheme independent. As a result, physical amplitude depends on the $\mu$ scale. However, in the naive factorization approach, the long-distance effects are disentangled from the short-distance effect assuming that the matrix element ${\langle{\cal O}\rangle}$ at the $\mu$ scale contains nonfactorizable contributions. This results in the cancellation of the $\mu$ dependence and scheme dependence of $C_i(\mu)$.
 \par We neglect here the so-called $W$ exchange and annihilation diagram, since in the limit $M_W \to \infty$ they are connected by Fiertz transformation and doubly suppressed by a kinematic factor of order $(\frac{m_i^2}{M_W^2})$ \cite{A52}. We also discard the color octet current which emerges after the Fiertz transformation of color singlet operators. Clearly, these currents violate factorization since they cannot provide transitions to vacuum states. Taking into account the Fiertz reordered contribution, the relevant coefficients are not $C_1(\mu)$ and $C_2(\mu)$ but the combination 
 \begin{equation}
 	a_{1,2}(\mu)=c_{1,2}(\mu)+\frac{1}{N_c}c_{2,1}(\mu).
 \end{equation}
 The factorization approximation, in general, works well in the description of two-body nonleptonic decays of heavy mesons in the limit of a large number of colors. Assuming a large $N_c$ limit to fix the QCD coefficients $a_1\approx c_1$ and $a_2\approx c_2$ at $\mu \approx m_b^2$, nonleptonic decays of heavy mesons have been analyzed in Refs. \cite{A15,A24,A63,A64,A65,A66,A67}. 
 \par The hadronic matrix element of the weak current $J_\mu$ are covariantly expanded in terms of weak form factors as
 \begin{eqnarray}
  {\langle V_1(\vec{k})|A_\mu|B_F(\vec{p})\rangle}&&=f(q^2)\epsilon_\mu^*+a_+(q^2)(\epsilon^*.p)(p+k)_\mu \nonumber \\ &&+a_-(q^2)(\epsilon^*.p)(p-k)_\mu,   
 \end{eqnarray}
\begin{equation}
 \resizebox{.9\hsize}{!}{$\langle V_1(\vec{k})|V_\mu|B_F(\vec{p})\rangle=ig(q^2)\epsilon_{\mu\nu\rho\sigma}\epsilon^{*\nu}(p+k)^\rho(p-k)^\sigma,$}   
\end{equation}
 where $\epsilon^*$ is the polarization of the vector meson $V_1$. $p$ and $k$ represent the four-momentum of the parent meson $B_F$ and daughter meson $V_1$, respectively. With the four-momentum transfer $q=p-k\equiv (E,0,0,|\vec{q}|)$ and mass $m_{V_1}$, the polarization of the daughter meson $V_1$ is taken in the form
 \begin{equation}
 \epsilon_\mu^\pm\equiv \frac{1}{\sqrt{2}}(0,\mp1,-i,0),\ \epsilon_\mu^L\equiv\frac{1}{m_{V_1}}(\vert\vec{q}\vert,0,0,E).
 \end{equation}
 The matrix element of the current $J^\mu$ between vacuum and vector-meson $V_2$ in the final state can be parametrized in terms of meson decay constant $f_{V_2}$ as
\begin{equation}
{\langle V_2|J^\mu|0\rangle}=\epsilon_{V_2}^{*\mu}f_{V_2}m_{V_2}.
\end{equation} 
In the factorization approach, the nonleptonic transition amplitude can be calculated from one of the three possible tree-level diagrams shown in Fig. 1. The color-favored transitions, shown in quark level diagram in Fig. 1(a), represent \textquotedblleft class I\textquotedblright \ transitions which are characterized by external emission of $W$-boson. In these transitions, the factorized amplitude coupled to the QCD factor $a_1$ only give the nonvanishing contribution. On the other hand, color-suppressed transitions shown in the diagram in Fig. 1(b) representing \textquotedblleft class II\textquotedblright \ transitions are characterized by internal $W$ emission. In such transitions, the nonvanishing contribution to the decay rate comes from factorized amplitude proportional to the QCD factor $a_2$. Figure 1(c), however, represents \textquotedblleft class III\textquotedblright \ transitions which are due to both color-favored and color-suppressed diagrams. In such decays, the factorized amplitudes corresponding to $a_1$ and $a_2$ contribute coherently to give the transition amplitude.\\
\par For the color-favored general type tree-level transition $B_F\to V_1(nS)V_2$ pertaining to \textquotedblleft class I\textquotedblright \ transitions, the decay rate can be written as \cite{A35,A36,A46}
\begin{equation}
	\Gamma={\frac{{ G_F}^2}{16\pi}}{a_1^2(\mu)}|V_{bq^{'}}V_{q_i\bar{q}_j}|^2{\frac{|\vec{k|}}{M^2}}{|\cal A|}^2,
\end{equation}

\begin{figure*}[!hbt]
	\centering
	\includegraphics[width=0.9\textwidth]{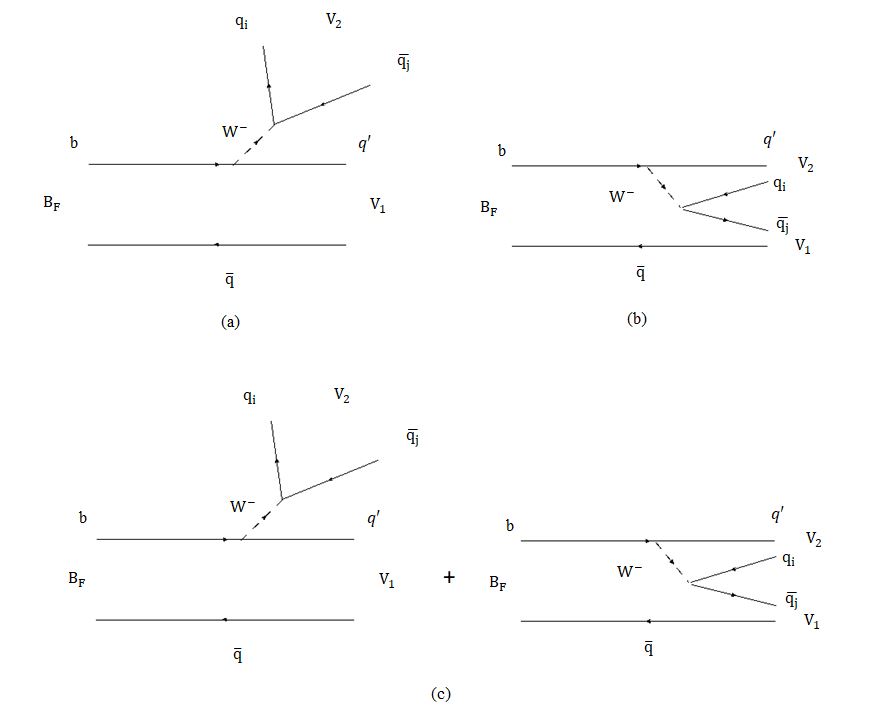}
\captionsetup[justification=centering]{}
	\caption { Quark-level diagram of the nonleptonic decay of a meson $B_F\to V_1(nS)V_2$.}
\end{figure*}
\noindent where $M$ and $\vec{k}$ represent the parent-meson mass and three-momentum of the recoiled daughter meson $V_1$, respectively, in the parent-meson rest frame. 
${|\cal A|}^2$ is the sum of the polarized amplitude squared with ${\cal A}_j \equiv {\langle V_2|J^\mu|0\rangle}{\langle V_1|J_\mu|B_F\rangle}$, such that

 \begin{equation}
 {|\cal A|}^2=\sum_{j}{|{\cal A}_j|^2}.
 \end{equation}
We use the notation $j=\pm$, $\mp$  or  $ll$, where the first and second labels denote the helicity of the $V_1$ and $V_2$ meson, respectively. From the polarized amplitudes expressed in terms of the weak form factors $f$, $g$ and $a_+$ and the decay constant $f_{V_2}$ shown in Eqs.(4)–(7), it is straightforward to find expressions for the positive, negative, and longitudinal polarizations, respectively, of the daughter meson $V_1$ as 
 \begin{eqnarray}
  {\cal A}_\pm&=&-f_{V_2}m_{V_2}\{{f(q^2)+2g(q^2)|\vec{k}|M}\},\nonumber
 \end{eqnarray} 
 \begin{eqnarray}
  {\cal A}_\mp&=&-f_{V_2}m_{V_2}\{{f(q^2)-2g(q^2)|\vec{k}|M}\},\nonumber
 \end{eqnarray}
\begin{eqnarray}
 {\cal A}_{ll}&&=\frac{f_{V_2}}{m_{V_1}}\Big[f(q^2)\Big
\{|\vec{k}|^2+\frac{1}{4M^2} (M^2+m_{V_1}^2-m_{V_2}^2) \nonumber\\ && \times (M^2+m_{V_2}^2-m_{V_1}^2)\Big\}+2a_+(q^2)|\vec{k}|^2M^2\Big],   
\end{eqnarray}
where
\begin{equation}
	|\vec{k}|=\Big[\Big(\frac{M^2+m_{V_1}^2-m_{V_2}^2}{2M}\Big)^2-m_{V_1}^2\Big]^{1/2}.
\end{equation} 
The decay widths and branching fractions for $B_F\to V_1(nS)V_2$ decays can be predicted from Eq.(8) using the expressions in Eqs. (9–11) for the polarized amplitudes in terms of the weak form factors derivable in the framework of the RIQ model.
\section{Transition amplitude and weak form factors}
As discussed in the preceding section, the nonleptonic transition amplitude for the process $B_F\to V_1(nS)V_2$ can be calculated from the tree-level diagram shown in Fig.1. The class-I type decay modes, depicted in Fig 1(a), are induced by the $b$-quark transition to the daughter quark $q^{'}$ with the emission of $W$-boson. The daughter quark $q^{'}$ and the antiquark $\bar{q}$ of the decaying parent meson state $|B_F(\vec{p},S_{B_F}) \rangle$ hadronize to form a vector meson state $|V_1(\vec{k},S_{V_1}) \rangle$. The externally emitted $W$-boson first decays to a quark-antiquark  pair ($q_i$ ${\bar{q}_j}$), which subsequently hadronizes to other vector meson state $|V_2(\vec{q},S_{V_2}) \rangle$.\\
 \par The decay process, in fact, occurs physically in the momentum eigenstate of participating mesons. Therefore, a field-theoretic description of a decay process demands meson-bound states to be represented by appropriate momentum wave packets reflecting momentum and spin distribution between the quark constituents in the meson core. A brief discussion of the wave-packet representation of meson bound state in the RIQ model is given in the Appendix. 
 Using the wave-packet representation (A.9-A.11) of participating meson states, the residual dynamics responsible for the decay process can, therefore, be described at the constituent level by the otherwise unbound quark and antiquark using the usual Feynman technique. The constituent-level $S$-matrix element $S_{fi}^{b\to q^{'}{q_i}{\bar{q}_j}}$ obtained from the appropriate Feynman diagram when operated upon by the bag-like operator ${\hat{\Lambda}}(\vec{p},S_{B_F})$ in the wave packet representation can give rise to the mesonic-level $S$-matrix element in the form
\begin{equation}
	S_{fi}^{B_F\to V_1(nS)V_2}\to \hat{\Lambda}(\vec{p},S_{B_F}) \ S_{fi}^{b\to q^{'}q_i {\bar{q}_j}}.
\end{equation}
\par Using the wave packet representation of the parent and daughter meson state, $|B_F(\vec{p},S_{B_F})\rangle$ and $|V_1(\vec{k},S_{V_1})\rangle$, respectively, we calculate the Feynman Diagram Fig. 1(a) and obtain the $S$-matrix element in the parent meson rest frame in the general form:
\begin{equation}
 \resizebox{.9\hsize}{!}{$S_{fi}=(2\pi)^4 \ {\delta^{(4)}(M-q-k)} \ (-i {\cal M}_{fi}){\frac{1}{\sqrt{V2M}}}\prod_{f}{\frac{1}{\sqrt{V2E_f}}}.$}   
\end{equation}
where the invariant transition amplitude ${\cal M}_{fi}$ is obtained in the form:
\begin{equation}
	{\cal M}_{fi}=\frac{G_F}{\sqrt{2}} {V_{bq^{'}} V_{q_i\bar{q}_j}} a_1  {\cal A},
\end{equation}
with ${\cal A} =h^\mu H_\mu$, $h^\mu=\epsilon_{V_2}^{*\mu}(\vec{q},\lambda_2)f_{V_2}m_{V_2}$ and

\begin{eqnarray}
H_\mu=&&\frac{1}{\sqrt{N_{B_F}(0)N_{V_1}(\vec{k})}}\int d^3\vec{p}_b \nonumber  \\ && \frac{{\cal G}_{B_F}(\vec{p}_b,-\vec{p}_b) \ {\cal G}_{V_1}(\vec{p}_b+\vec{k},-\vec{p}_b)}{\sqrt{E_b(\vec{p}_b)E_{q^{'}}(\vec{p}_b+\vec{k})}}\nonumber \\ && \times \sqrt{[E_b(\vec{p}_b)+E_{\bar{q}}(-\vec{p}_b)][E_{q^{'}}(\vec{p}_b+\vec{k})+E_{\bar{q}}(-\vec{p}_b)]}\nonumber \\ && \langle S_{V_1}|J_\mu(0)|S_{B_F}\rangle.
\end{eqnarray} 
The terms $E_b(\vec{p}_b)$ and $E_{q^{'}}(\vec{p}_b+\vec{k})$ in (15) stand for the energy of the non-spectator quark of the parent and daughter meson, $\vec{p}_b$ and $\vec{k}$ represent three momentum of the non-spectator constituent quark $b$ and the daughter meson $V_1$, respectively and $q=p-k$ is the four-momentum transfer. Finally, $\langle S_V|J_\mu|S_{B_F}\rangle$ is the symbolical representation of the spin matrix elements of the  effective vector-axial vector current; which can be written in the explicit form:
\begin{eqnarray}
	\langle S_{V_1}|J_\mu(0)|S_{B_F}\rangle&&=
	\sum_{{\lambda_b},{\lambda_{q^{'}}},{\lambda_{\bar{q}}}}\zeta_{b\bar{q}}^{B_F}(\lambda_b,\lambda_{\bar{q}}) \ \zeta_{q^{'}\bar{q}}^{V_1}(\lambda_{q^{'}},\lambda_{\bar{q}}) \nonumber \\ && \bar{u}_{q^{'}}(\vec{p}_b+\vec{k},\lambda_{q^{'}}) \ \gamma_\mu(1-\gamma_5) \ u_b(\vec{p}_b,\lambda_b).
\end{eqnarray}
Here, $u_i$ stands for free Dirac spinor.$\zeta^{B_F}(\lambda_b,\lambda_{\bar{q}})$ and $\zeta^{V_1}(\lambda_{q^{'}},\\ \lambda_{\bar{q}})$ are the appropriate $SU(6)$ spin flavor coefficients corresponding to the parent and daughter meson, respectively.
\par It may be pointed out here that, in our description of the decay process, $B_F\to V_1(nS)V_2$ the three momentum conservation is ensured explicitly via $\delta^{(3)}(\vec{p}_b+\vec{p}_{\bar{q}}-\vec{p})$ and $\delta^{(3)}(\vec{p}_i\\+\vec{p}_j-\vec{k})$ in the participating meson states. However, energy conservation in such a scheme is not ensured so explicitly. This is in fact a  typical problem in all potential model descriptions of mesons as bound states of valence quarks and antiquarks interacting via some instantaneous potential. This problem has been addressed in the previous analysis in this model in the context of radiative leptonic decays of heavy flavored meson $B$, $B_c$, $D$, $D_s$ \cite{A68,A69,A70} and also in the QCD relativistic quark model \cite{A71,A72}, where the effective momentum distribution function ${\cal G}_{B_F}(\vec{p}_b,\vec{p}_{\bar{q}})$ that embodies bound-state characteristics of the meson ensures energy conservation in an average sense satisfying $E_M=\langle B_F(\vec{p},S_{B_F})|[E_b(\vec{p}_b)+E_{\bar{q}}(\vec{p}_{\bar{q}})]|B_F(\vec{p},S_{B_F}) \rangle$. In view of this, we take the energy conservation constraint $M=E_b(\vec{p}_b)+E_{\bar{q}}(-\vec{p}_b)$ in the parent meson rest frame. This along with the three momentum conservation via appropriate $\delta^{(3)}(\vec{p}_b+\vec{p}_{\bar{q}}-\vec{p})$ in the meson state ensures the required four-momentum  conservation ${\delta^{(4)}(p-k-q)}$ at the mesonic level, which is pulled out of the quark-level integration to obtain the $S$-matrix element in the standard form (13). This has been discussed elaborately in earlier works \cite{A33,A34,A35,A36,A55}. 
\par Using usual spin algebra the spacelike and timelike components of the spin matrix elements $\langle S_{V_1}|J_\mu(0)|S_{B_F}(0)\rangle$ corresponding to vector and axial vector current are obtained in the form 
\begin{equation}
 \langle S_{V_1}(\vec{k},\hat{\epsilon}_{V_1}^*)|V_0|S_{B_F}(0)\rangle=
	0,   
\end{equation}
 \begin{equation}
     \resizebox{.9\hsize}{!}{$\langle S_{V_1}(\vec{k},\hat{\epsilon}_{V_1}^*)|V_i|S_{B_F}(0)\rangle=\frac{i[E_b(\vec{p}_b)+m_b]}{\sqrt{[E_b(\vec{p}_b)+m_b][E_{q^{'}}(\vec{p}_b+\vec{k})+m_{q^{'}}]}}(\hat{\epsilon}_{V_1}^*\times \vec{k})_i$,}
 \end{equation}
 \begin{equation}
     \resizebox{.9\hsize}{!}{$\langle S_{V_1}(\vec{k},\hat{\epsilon}_{V_1}^*)|A_i|S_{B_F}(0)\rangle=\frac{[E_b(\vec{p}_b)+m_b][E_{q^{'}}(\vec{p}_b+\vec{k})+m_{q^{'}}]-(\vec{p}_b^2/3)}{\sqrt{[E_b(\vec{p}_b)+m_b][E_{q^{'}}(\vec{p}_b+\vec{k})+m_{q^{'}}]}}(\hat{\epsilon}_{V_1}^*)_i$,}
 \end{equation}
\begin{equation}
     \resizebox{.9\hsize}{!}{$\langle S_{V_1}(\vec{k},\hat{\epsilon}_{V_1}^*)|A_0|S_{B_F}(0)\rangle=\frac{[E_b(\vec{p}_b)+m_b]}{\sqrt{[E_b(\vec{p}_b)+m_b][E_{q^{'}}(\vec{p}_b+\vec{k})+m_{q^{'}}]}}(\hat{\epsilon}_{V_1}^*.\vec{k})$.}
 \end{equation}		
Here $E_b(\vec{p}_b)=\sqrt{\vec{p}_b^2+m_b^2}$ and $E_{q^{'}}(\vec{p}_b+\vec{k}) =\sqrt{(\vec{p}_b+\vec{k})^2+m_{q^{'}}^2}$ are, respectively, the energy of the non-spectator quark $b$ and daughter quark $q^{'}$. The spacelike component of the hadronic matrix element $H_\mu$ obtained from Eq.(15) via Eqs. (18) and (19) are compared with the corresponding expressions from Eqs. (4) and (5), which lead to the model expressions of the weak form factors $g(q^2)$ and $f(q^2)$ in the form:
\begin{eqnarray}
g(q^2)&=&-\frac{1}{2M}\int d\vec{p}_b \ {\cal Q}(\vec{p}_b) \ [E_b(\vec{p}_b)+m_b],\\
f(q^2)&=&-\int d\vec{p}_b{\cal R}(\vec{p}_b),
\end{eqnarray}
where
\begin{eqnarray}
	{\cal Q}(\vec{p}_b)&&=\frac{{\cal G}_{B_F}(\vec{p}_b,-\vec{p}_b){\cal G}_{V_1}(\vec{p}_b+\vec{k},-\vec{p}_b)}{\sqrt{N_{B_F}(0)N_{V_1}(\vec{k})}}  
	 \nonumber \\ && \frac{\sqrt{[E_b(\vec{p}_b)+E_{\bar{q}}(-\vec{p}_b))][E_{q^{'}}(\vec{p}_b+\vec{k})+E_{\bar{q}}(-\vec{p}_b)]}}{\sqrt{E_b(\vec{p}_b)E_{q^{'}}(\vec{p}_b)[E_b(\vec{p}_b)+m_b][E_{q^{'}}(\vec{p}_{q^{'}})+m_{q^{'}}]}},
	\end{eqnarray}
\begin{equation}
{\cal R}(\vec{p}_b)={\cal Q}(\vec{p}_b)\Big[(E_b(\vec{p}_b)+m_b)(E_{q^{'}}(\vec{p}_{q^{'}})+m_{q^{'}})-\frac{\vec{p}_b^2}{3}\Big].
\end{equation}	
The timelike component of hadronic amplitude obtained from Eq. (15) via Eq. (20), when compared with the corresponding expression from Eq. (4) yields an expression of the form factor $a_+(q^2)$ in the form: 
\begin{eqnarray}
	a_+(q^2)&&=\frac{1}{2M^2}\Big[E_{V_1}\int d\vec{p}_b \ {\cal Q}(\vec{p}_b) \ (E_b(\vec{p}_b)+m_b)-\nonumber \\ &&\int d\vec{p}_b \ {\cal R}(\vec{p}_b)\Big].
\end{eqnarray}
Then it is straightforward to get the model expression for the polarized amplitude squared $|{\cal A}_j|^2$ using Eqs.(21-25). Summing over possible polarization states and integrating over the final-state particle momenta, the decay width $\Gamma(B_F\to V_1(nS)V_2)$ is obtained in the parent-meson rest frame from the generic expression
\begin{eqnarray}
	\Gamma(B_F\to V_1(nS)V_2)&&=\frac{1}{(2\pi)^2}\int \frac{d\vec{k} \ d\vec{q}}{2M2E_{V_1}2E_{V_2}}\nonumber \\ &&\delta^{(4)}(p-k-q)\times \sum |{\cal M}_{fi}|^2.
\end{eqnarray}  
\par The two-body nonleptonic decay $(B_F\to V_1(nS)V_2)$, described so far in this section refers to the color-favored \textquotedblleft class I\textquotedblright\ decays involving external emission of $W$-boson. Similarly, class II and III type $B_F\to V_1(nS)V_2$ decays can be calculated from the corresponding Feynman diagrams shown in Figs. 1(b) and 1(c), respectively. The model expressions for relevant form factors and decay rates for such decays (class II and class III) can be obtained by suitable replacement of appropriate flavor degree of freedom, quark masses, quark binding energies, QCD factors $a_1, a_2$, and the meson decay constants. 
 \section{Numerical results and Discussion}
 In this section, we present our numerical results in comparison with other model predictions and the available experimental data. For numerical calculation, we use the model parameters $(a,V_0)$, quark mass $m_q$ and quark binding energies $E_q$, which have been fixed from hadron spectroscopy by fitting the data of heavy and heavy-light flavored mesons in their ground state as \cite{A73,A74,A75}
 \begin{eqnarray}
 	(a,V_0)&=&(0.017166 \ GeV^3, -0.1375 \ GeV),\nonumber\\
  (m_b, m_c)&=&(4.77659, 1.49276) \ GeV,\nonumber\\
  (m_s, m_u=m_d)&=&(0.31575, 0.07875) \ GeV,\nonumber\\
   (E_b, E_c)&=&(4.76633, 1.57951) \ GeV,\nonumber \\ 
   (E_s, E_u=E_d)&=&(0.591,0.47125) \ GeV.   
 \end{eqnarray}  
The description of the decay process involving radially excited meson states, the constituent quarks in the meson-bound states are expected to have higher binding energies compared to their ground-state binding energies. For this, we solve the cubic equation representing the binding energy condition (A.5) for respective constituent quarks $(c, s, u=d)$ in radially excited $2S$ and $3S$ states of the $(\bar{c}c)$, $(\bar{c}u)$, $(\bar{c}d)$ systems as 
\begin{eqnarray}
  (E_c, E_s, E_u=E_d)_{2S}&=&(1.97015, 1.07737, 0.96221)GeV,\nonumber\\
  (E_c, E_s, E_u=E_d)_{3S}&=&(2.22478, 1.40043, 1.29359)GeV.
\end{eqnarray}
Using the above input parameters (27), wide-ranging hadronic phenomena have been described within the framework of the RIQ model, which includes the two body nonleptonic decays of $B$ and $B_c$ mesons to ground state mesons in the charmonium, charm, strange and non-strange light flavor sectors \cite{A33,A34,A35,A36}. For CKM parameters and the lifetime of decaying mesons, we take their respective central values from the Particle Data Group (2022) \cite{A82}:
 \begin{eqnarray}
 	(V_{bc}, V_{bu})&=&(0.0408, 0.00382),\nonumber\\
 (V_{cs}, V_{cd})&=&(0.975, 0.221),\nonumber\\
 	(V_{us}, V_{ud})&=&(0.2243, 0.97373),
 \end{eqnarray}
and
\begin{eqnarray}
	(\tau_{B_c^-}, \tau_{B^-})&=&(0.510, 1.638) \ ps, \nonumber\\
	(\tau_{\bar{B^0}}, \tau_{\bar{B_s^0}})&=&(1.519, 1.5672) \ ps,
\end{eqnarray}
respectively. For the mass and decay constant of the participating mesons, considered as phenomenological inputs in the numerical calculation, we take the central values of the available observed data from Ref. \cite{A5,A6,A82}. In the absence of the observed data on the mass of excited ($2S$ and $3S$) charmed and strange-charmed mesons and the meson decay constants, we take the corresponding predicted data from established theoretical approaches \cite{A83,A84,A85,A86}. Accordingly, the updated meson masses and decay constants used in the present study are listed in Table-1.
 \begin{table}[!hbt]
 	\renewcommand{\arraystretch}{1.3}
 	\centering
 	\setlength\tabcolsep{7pt}
 	\caption{The masses and decay constants of mesons.}
 	\label{tab1}
 	\begin{tabular}{lllll}
 		\hline Particle& &\  Mass(MeV) &\ \ Decay constant(MeV)&  \\
 		\hline
 		$\rho^{-}$&&\ 775.11\ \ \cite{A82}&\ \ \ \ \ \ \ \ \ 208.5\cite{A84}&\\
 		$K^{*\pm}$&&\ 891.67\ \ \cite{A82}&\ \ \ \ \ \ \ \ \ 217 \ \ \cite{A85}&\\
 		$D^{*0}(1S)$&&\ 2006.8\ \ \cite{A82}&\ \ \ \ \ \ \ \ \ 339 \ \ \cite{A86}&\\
 		$D^{ *\pm}(1S)$&&\ 2010.2\ \ \cite{A82}&\ \ \ \ \ \ \ \ \ 341 \ \ \cite{A86} & \\
 		$D_s^{*\pm }(1S)$& &\ 2112.2\ \ \cite{A82}&\ \ \ \ \ \ \ \ \ 375 \ \ \cite{A86}& \\
 		$J/\psi (1S)$& &\  3096.9\ \ \cite{A82} &\ \ \ \ \ \ \ \ \ 459 \ \ \cite{A86}&\\
 		$B_c^-$& &\ 6274.47\cite{A82}& &\\
 		$\bar{B}^0$& &\ 5279.6\ \ \cite{A82}& &\\
 		$\bar{B}_s^0$& &\ 5415.4\ \ \cite{A82}& &\\
 		$B^-$& &\ 5279.3\ \ \cite{A82}& &\\
 		$D^{*\pm}(2S)$&&\ 2637\ \ \ \ \ \cite{A5}&\ \ \ \ \ \ \ \ \ 290 \ \ \cite{A86}&\\
 		$D^{*0}(2S)$ &&\ 2681\ \ \ \ \  \cite{A6}&\ \ \ \ \ \ \ \ \ 289 \ \ \cite{A86}&\\
 		$D_s^{*\pm}(2S)$&&\ 2732\ \ \ \ \ \cite{A83}&\ \ \ \ \ \ \ \ \ 312 \ \ \cite{A86}&\\
 		$\psi(2S)$&& \ 3686.1\ \ \cite{A82}&\ \ \ \ \ \ \ \ \ 364 \ \ \cite{A86}&\\		
 		$D^{*0}(3S)$&&\ 3080\ \ \ \ \ \cite{A18}&&\\
 		$D^{*\pm}(3S)$&&\ 3110\ \ \ \ \ \cite{A83}&&\\
 		$D_s^{*\pm}(3S)$&&\ 3193\ \ \ \ \ \cite{A83}&&\\
 		$\psi(3S)$&&\ 4039.1\ \  \cite{A82}&\ \ \ \ \ \ \ 
 \ \ 319 \ \ \cite{A86}&\\
 		
 		\hline
 	\end{tabular}
 \end{table}
\par It may be mentioned here that, in the prediction of nonleptonic decay, uncertainties mostly creep into the calculation through input parameters: potential parameter $(a, V_0)$, quark mass $(m_q)$ and quark binding energy $(E_q)$, CKM parameters, meson decay constants and QCD coefficients \\ $(a_1, a_2)$ etc. As mentioned above, the potential parameters, quark masses and quark binding energies (27,28) have already been fixed at the static level application of RIQ model by fitting the mass spectra of heavy and heavy-light mesons. In order to avoid uncertainty in our model predictions, we take the central values of the CKM parameter as well as the observed value of the decay constants. As such we do not have the liberty to use any free parameter in our calculation which could be fine-tuned from time to time to predict any hadronic phenomena. In that sense, we perform almost a parameter-free calculation in our studies. As regards the QCD coefficients: ($a_1$, $a_2$), different sets of data for decays induced by the $b$-quark transition at the quark level, are used in the literature. For example, Colangelo and De Fazio, in Ref. \cite{A27} use QCD coefficients, Set 1: $(a_1^b, a_2^b)=(1.12,-0.26)$, as fixed in Refs. \cite{A87,A88}. In most earlier calculations, the authors use a different set of QCD coefficients, Set 2: $(a_1^b, a_2^b)=(1.14, -0.2)$, fixed by Buras {\it et al.} \cite{A52} in the mid-1980s, whereas Dubnicka {\it et al.} \cite{A89} use a different set of numerical value, i.e., Set 3: $(a_1^b, a_2^b)=(0.93,-0.27)$. We use all three sets of the Wilson coefficients in our calculation.
\par Before using the above input parameters in our numerical analysis, it is pertinent to elaborate a bit on the energy conservation ansatz mentioned in Sec.3. The present analysis based on the energy conservation constraints $M=E_b(\vec{p}_b)+E_{q^{'}}(-\vec{p}_b)$ in the parent meson rest frame might lead to spurious kinematic singularities at the quark-level integration appearing in the decay amplitude. This problem has already been addressed previously in their QCD relativistic quark model approach \cite{A71,A72} and later by our group in the study of radiative leptonic decays of heavy and heavy-light flavored meson sector \cite{A68,A69,A70}, by assigning a running mass $m_b$ to the non-spectator quark that satisfies the relation: 
\begin{eqnarray}
 m_b^2(|\vec{p}_b|)&=&M^2-m_{\bar{q}}^2-2M\sqrt{|\vec{p}_b|^2+m_{\bar{q}}^2},\nonumber   
\end{eqnarray}
as an outcome of the energy conservation ansatz, while retaining definite mass $m_{\bar{q}}$ of the spectator quark $\bar{q}$. This leads to an upper bound on the quark momentum $|\vec{p}_b|<\frac{M^2-m_{\bar{q}}^2}{2M}$  in order to retain $m_b^2(|\vec{p}_b|)$ the positive definite. The upper limit $|\vec{p}_b|_{max}$ would have no other bearing to seriously affect the calculation which is apparent from the shape of the radial quark momentum distribution $|\vec{p}_b|{\cal G}(\vec{p}_b,-\vec{p}_b)$. In fact, the quark momentum distribution obtained in this model \cite{A68,A69,A70} is similar to the prediction of the QCD relativistic quark model analysis \cite{A71,A72}. The rms value of the active quark momentum $\sqrt{\langle |\vec{p}_b^2|\rangle}$, where $\langle |\vec{p}_b^2|\rangle=\langle B_F(0)|\vec{p}_b^2|B_F(0)\rangle$, the expectation value of the binding energies of the active quark $b$, and spectator $\bar{q}$ and the sum of the binding energy of quark and antiquark pair $\langle E_b(\vec{p}_b^2)\rangle$, $\langle E_{\bar{q}}(|-\vec{p}_b^2|)\rangle$ and $\langle[E_b(\vec{p}_b^2)+E_{\bar{q}}(|-\vec{p}_b^2|)] \rangle$, respectively, calculated in the framework of RIQ model, are presented in Table 2.
\begin{table*}[!hbt]
	\renewcommand{\arraystretch}{0.75}
	\centering
	\setlength\tabcolsep{0.25pt}
	\caption{The rms values of quark momentum, expectation values of the momentum of quark and antiquark, and the sum of the momentum of quark and antiquark in the meson states.}
	\label{tab2}
	\resizebox{\textwidth}{!}
	{
		\begin{tabular}{llllll}
			
			 \hline {\tiny Meson state} &\  {\tiny$\sqrt{\langle |\vec{p}_b^2|\rangle}$}&\  {\tiny$\langle E_b(\vec{p}_b^2)\rangle$}&\ {\tiny$\langle E_{\bar{q}}(|-\vec{p}_b^2|)\rangle$}&\  {\tiny$\langle[E_b(\vec{p}_b^2)+E_{\bar{q}}(|-\vec{p}_b^2|)]\rangle$}&\  {\tiny Observed meson }\\
			{\tiny $|X(0)\rangle$}&\ \ {\tiny (GeV)}&\ \ {\tiny (GeV)}&\ \ \ \ {\tiny (GeV)}&\ \ \ \ \ \ \ \ \ {\tiny (GeV)}&\ \ {\tiny mass (GeV)}\\
			\hline
			{\tiny$|B_u(0)\rangle$}&\ \ {\tiny 0.51}&\ \ {\tiny 4.799}&\ \ \ \ \  {\tiny 0.480}&\ \ \ \ \ \ \ \ \ \  {\tiny 5.279}&\ \ \ \ {\tiny 5.27925}\\
			{\tiny $|B_c(0)\rangle$}&\ \ {\tiny 0.66}&\ \ {\tiny 4.657}&\ \ \ \ \  {\tiny 1.629}&\ \ \ \ \ \ \ \ \ \  {\tiny 6.286}&\ \ \ \ {\tiny 6.27447}\\
			{\tiny $|D(0)\rangle$}&\ \ {\tiny 0.4506}&\ \ {\tiny 1.4418}&\ \ \ \ \  {\tiny 0.4275}&\ \ \ \ \ \ \ \ \ \   {\tiny 1.8693}&\ \ \ \ {\tiny 1.86965}\\
			{\tiny $|D_s(0)\rangle$}&\ \ {\tiny 0.4736}&\ \ {\tiny 1.4165}&\ \ \ \ \  {\tiny 0.5517}&\ \ \ \ \ \ \ \ \ \ {\tiny 1.9682}&\ \ \ \ {\tiny 1.96835}\\
			\hline
				\end{tabular}
        }
\end{table*}
\par It is noteworthy to discuss four important aspects of our present approach. (1) The rms value of the quark momentum in the meson-bound state is much less than the corresponding upper bound  $|\vec{p}_b|_{max}$, as expected. (2) The average energies of a constituent quark of the same flavor in different meson-bound states do not exactly match. This is because the kinematics and binding energy conditions for constituent quarks due to the color forces involved are different from one meson-bound state to other. The constituent quarks in the meson-bound state are considered to be free particles of definite momenta, each associated with its momentum probability amplitude derivable in this model via momentum space projection of the respective quark eigenmodes. On the other hand, the energies shown in Eq. (27,28), which are the energy eigenvalues of the corresponding bound quarks with no definite momenta of their own, are obtained from respective quark orbitals by solving the Dirac equation in this model. This makes a marginal difference between the energy eigenvalues (27,28) and the average energy of constituent quarks shown in Table 2. (3) The expectation values of the sum of the energy of a constituent quark and antiquark in the meson-bound state are obtained in good agreement with the corresponding observed meson masses as shown in Table 2. These important aspects of our results lend credence to our energy conservation ansatz in an average sense through the effective momentum distribution function like ${\cal G}_{B_F}(\vec{p}_b,-\vec{p}_b)$ in the meson-bound state $|B_F(0)\rangle$. This ansatz along with three momentum conservation in the meson-bound state (A.1) ensures the required energy-momentum conservation in our description of several decay processes pointed out earlier. In the absence of any rigorous field theoretic description of the meson-bound states, invoking such an ansatz is no doubt a reasonable approach for a constituent-level description of hadronic phenomena. (4) Finally, in a self-consistent dynamic approach, we extract the form factors from the overlapping integrals of meson wave functions, where the $q^2$ dependence of the decay amplitude is automatically encoded. This is in contrast to some model approaches cited in the literature where the form factors are determined only at one kinematic point, i.e., either at $q^2\to0$ or $q^2\to q^2_{max}$, and then extrapolated to the entire kinematic range using some phenomenological ansatz (mainly dipole or Gaussian form).\\
\par The invariant form factors $g(q^2)$, $f(q^2)$ and $a_+(q^2)$, which represent the decay amplitudes, are found to have different dimensions. In order to study their $q^2$-dependence over the allowed kinematic range, they need to be treated on equal footing by casting them in the dimensionless form :
\begin{eqnarray}
	V(q^2)&=&(M+m_{V_1})g(q^2),\nonumber\\
	A_1(q^2)&=&(M+m_{V_1})^{-1}f(q^2),\nonumber\\
	A_2(q^2)&=&-(M+m_{V_1}) a_+(q^2).
\end{eqnarray}
One may naively expect the form factor in the dimensionless form to satisfy the heavy quark-symmetry (HQS) relation:
\begin{equation}
	V(q^2)\simeq A_2(q^2)\simeq\tilde{A_1}(q^2),
\end{equation}
as an outcome of the heavy quark effective theory (HQET), where
\begin{equation}
\tilde{A_1}(q^2)=\Big[1-\frac{q^2}{(M+m_{V_1})^2}\Big]^{-1}A_1(q^2).
\end{equation}

\begin{figure}[!hbt]
	\centering
	\includegraphics[width=0.325\textwidth]{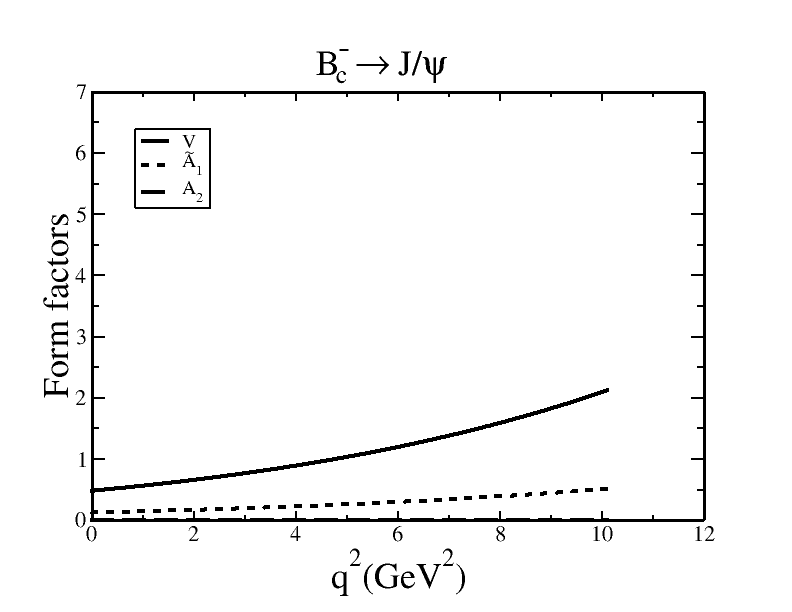}
	\includegraphics[width=0.325\textwidth]{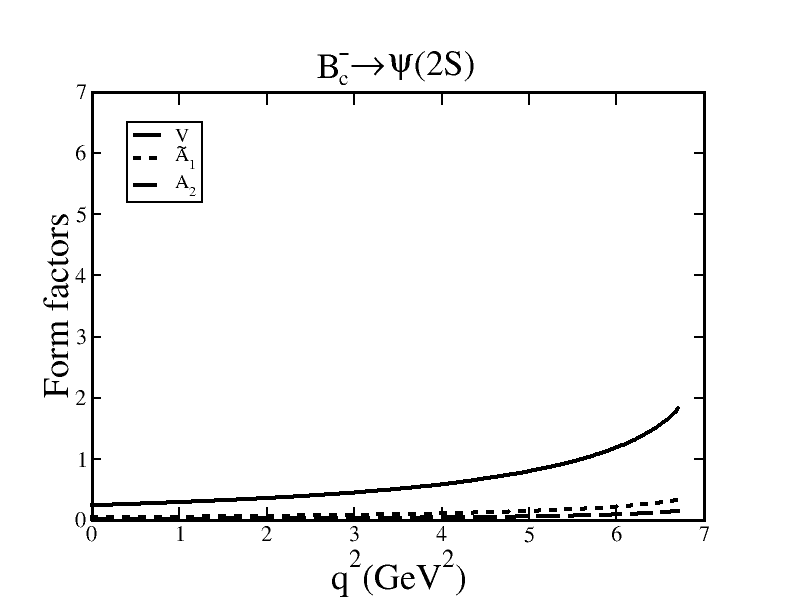}
	\includegraphics[width=0.325\textwidth]{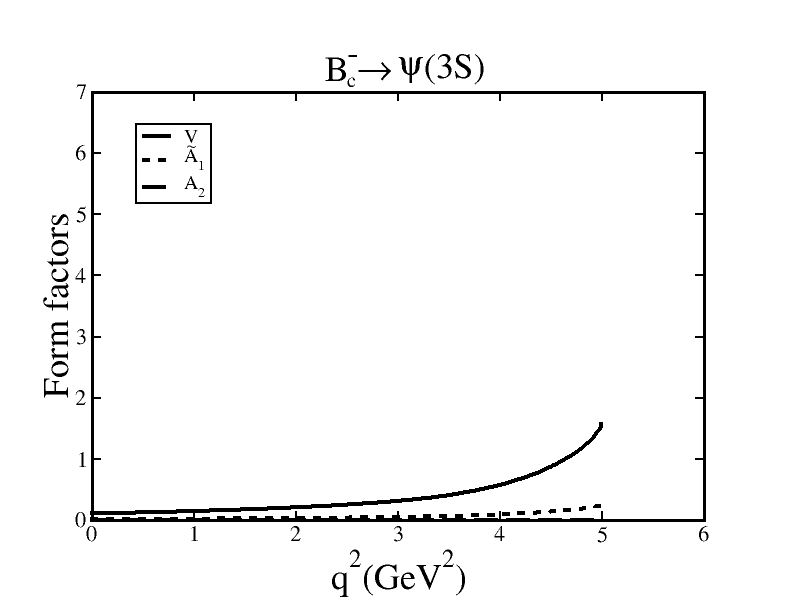}
	\caption{$q^2$-dependence of form factors in $B_c^-\to \psi(nS)$ type decays.}
\end{figure}
\begin{figure}[!hbt]
	\centering
	\includegraphics[width=0.325\textwidth]{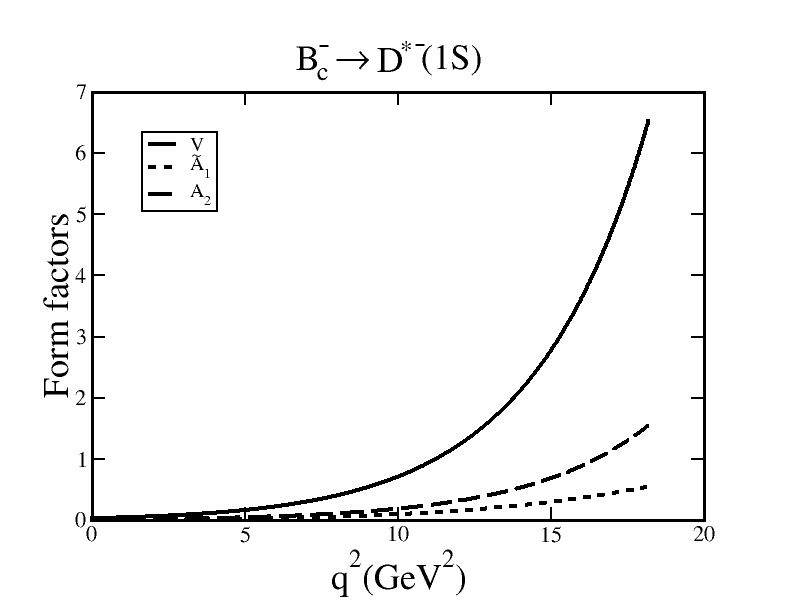}
	\includegraphics[width=0.325\textwidth]{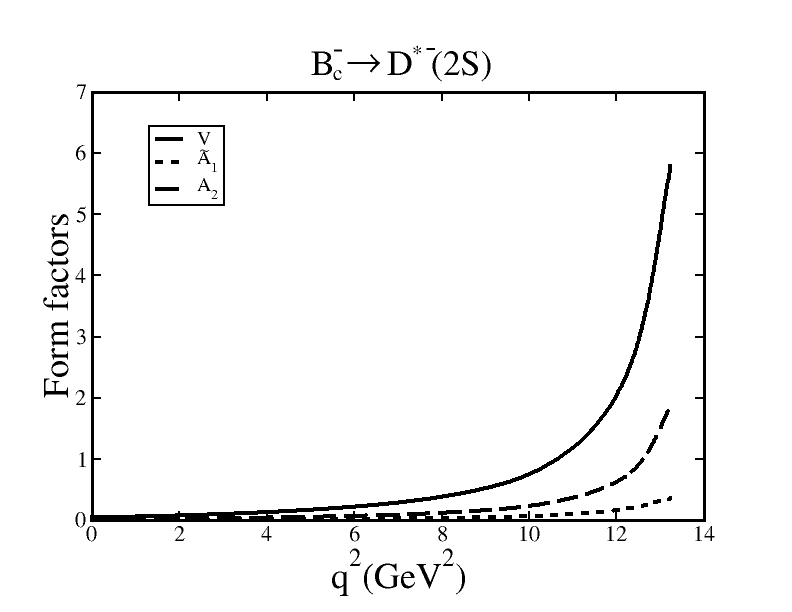}
	\includegraphics[width=0.325\textwidth]{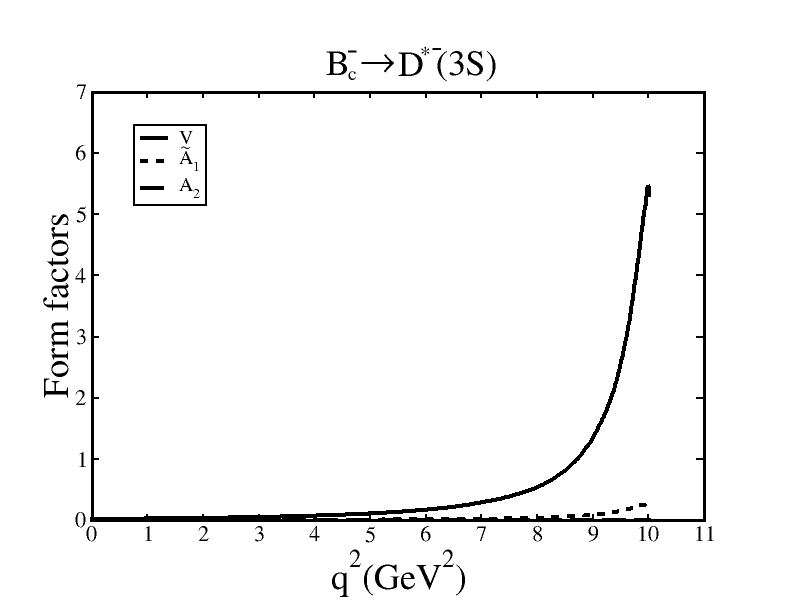}
	\caption{$q^2$-dependence of form factors in $B_c^-\to D^{*-}(nS)$ type decays.}
\end{figure}
\begin{figure}[!hbt]
	\centering
	\includegraphics[width=0.325\textwidth]{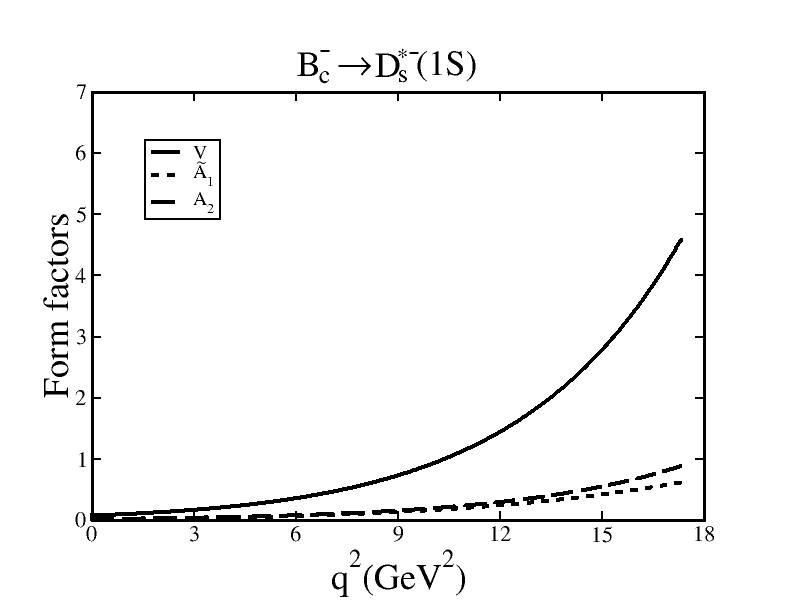}
	\includegraphics[width=0.325\textwidth]{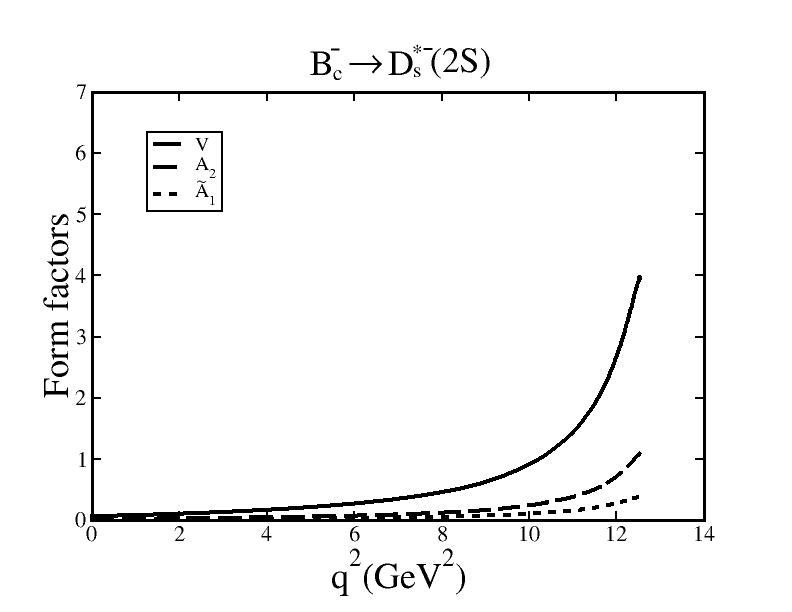}
	\includegraphics[width=0.325\textwidth]{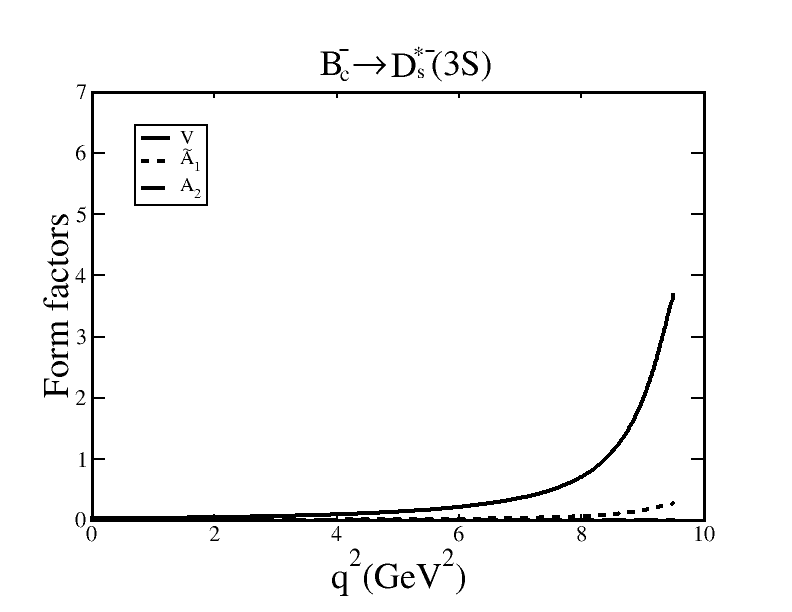}
	\caption{$q^2$-dependence of form factors in $B_c^-\to D_s^{*-}(nS)$ type decays.}
\end{figure}
\begin{figure}[!hbt]
	\centering
	\includegraphics[width=0.325\textwidth]{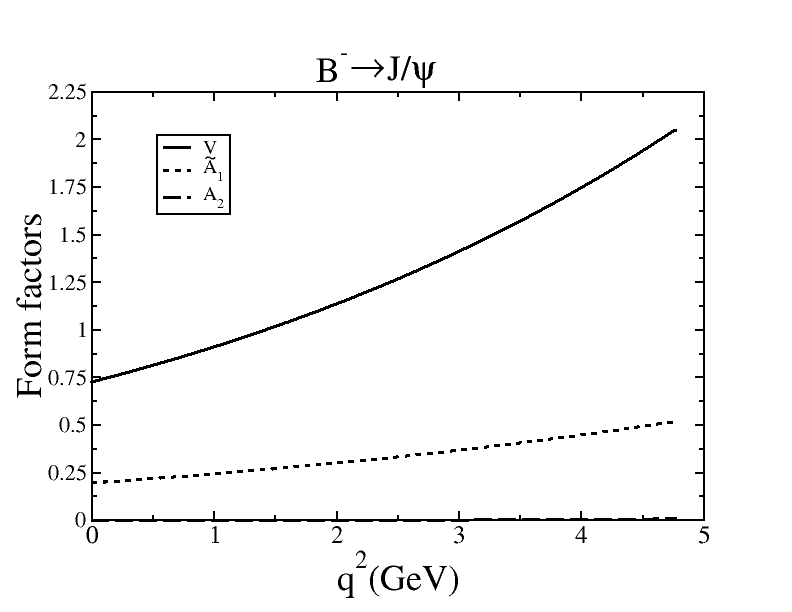}
	\includegraphics[width=0.325\textwidth]{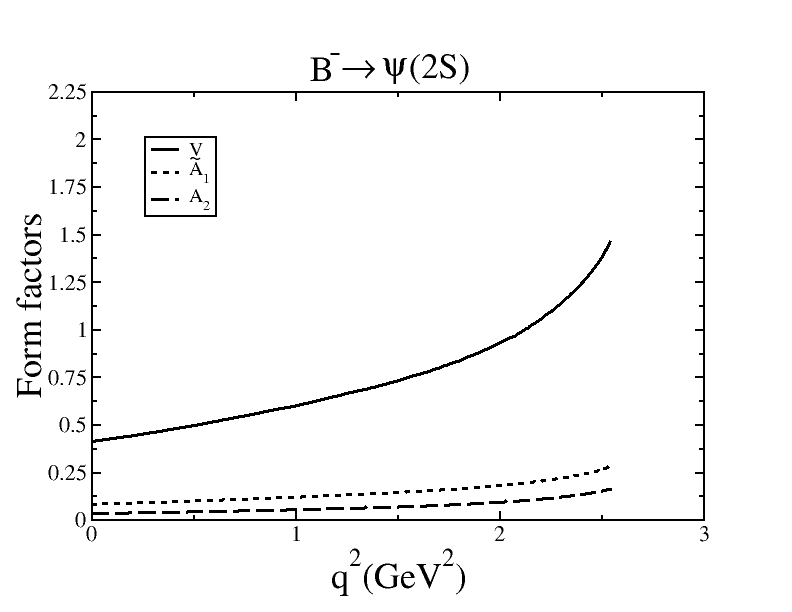}
	\includegraphics[width=0.325\textwidth]{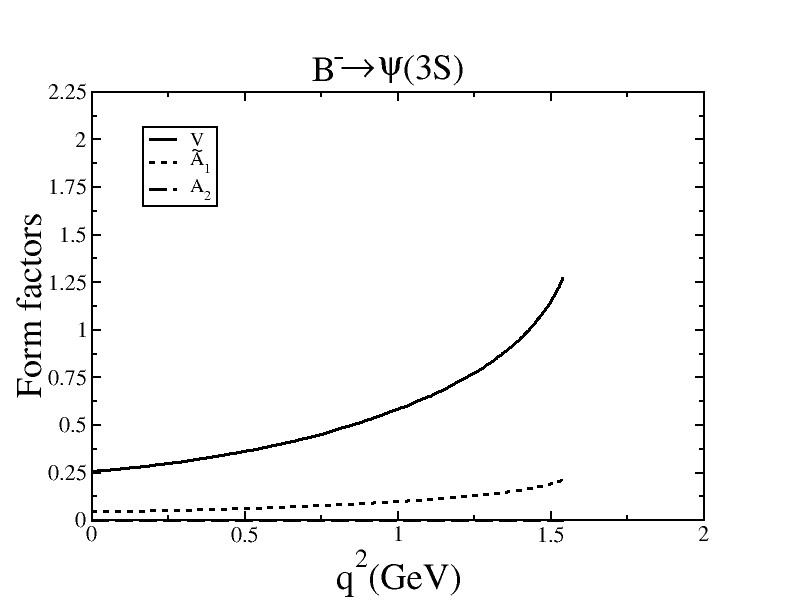}
	\caption{ $q^2$-dependence of form factors in $B^-\to \psi(nS)$ type decays.}
\end{figure}
\begin{figure}[!hbt]
	\centering
	\includegraphics[width=0.325\textwidth]{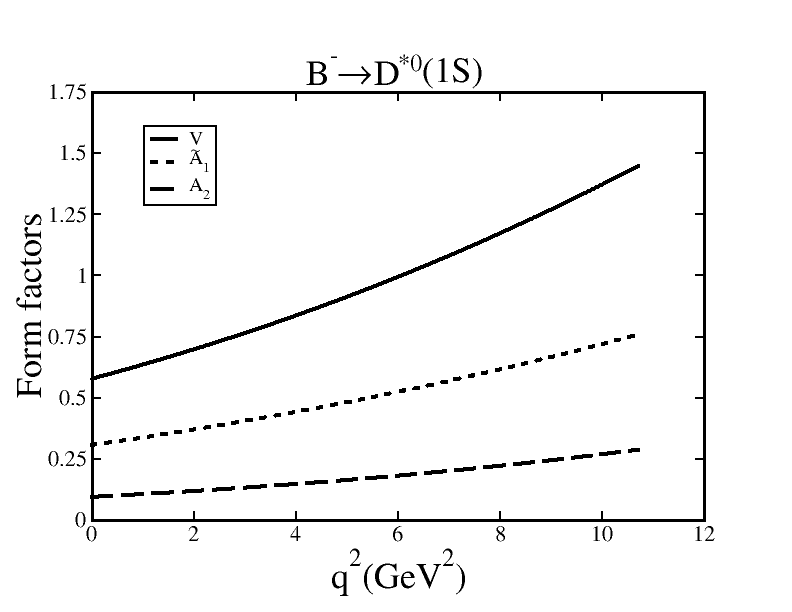}
	\includegraphics[width=0.325\textwidth]{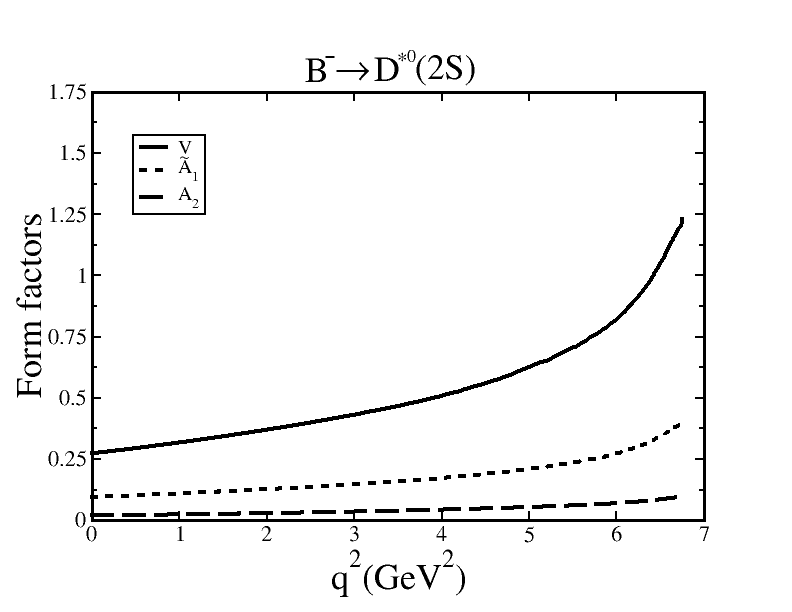}
	\includegraphics[width=0.325\textwidth]{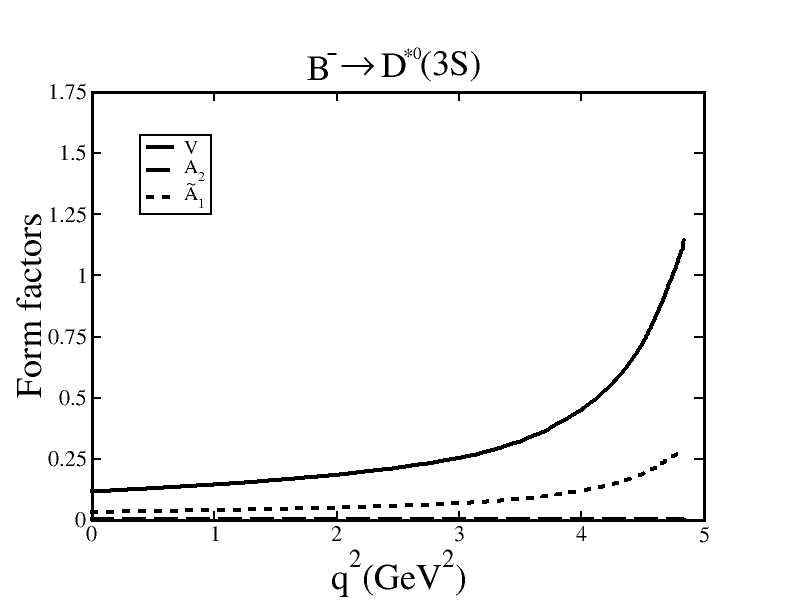}
	\caption{ $q^2$-dependence of form factors in $B^-\to D^{*0}(nS)$ type decays.}
\end{figure}
\begin{figure}[!hbt]
	\centering
	\includegraphics[width=0.325\textwidth]{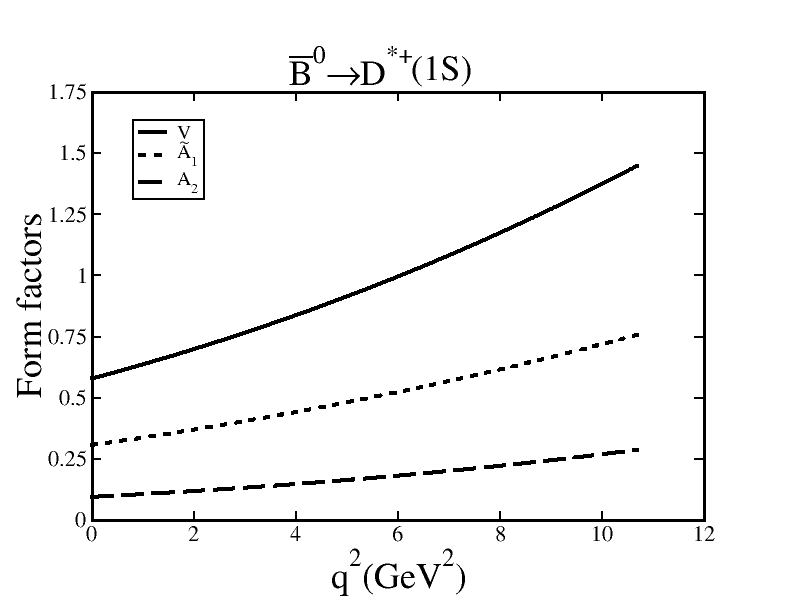}
	\includegraphics[width=0.325\textwidth]{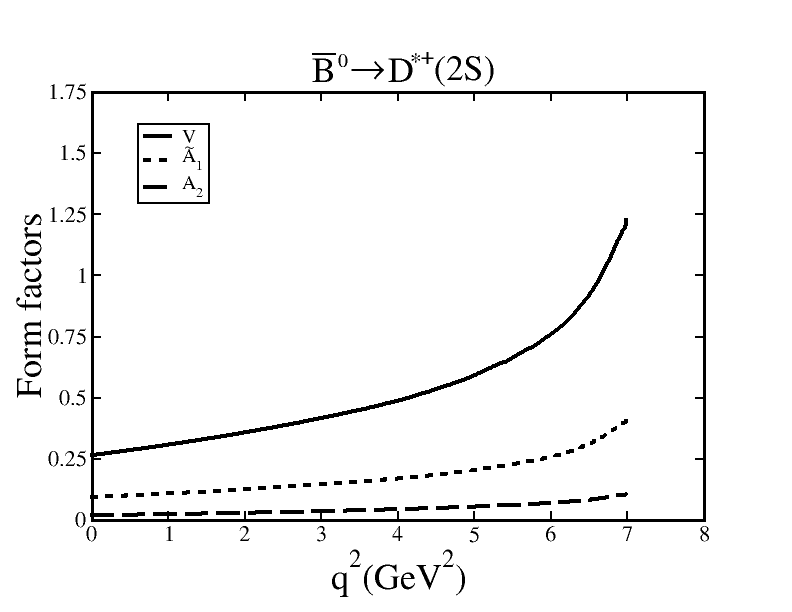}
	\includegraphics[width=0.325\textwidth]{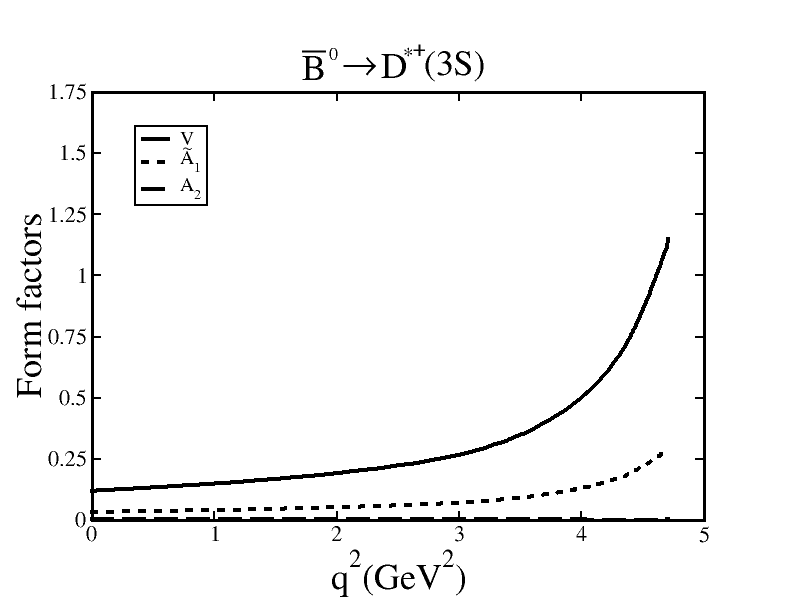}
	\caption{ $q^2$-dependence of form factors in $\bar{B^0}\to D^{*+}(nS)$ type decays.}
\end{figure}
\begin{figure}[!hbt]
	\centering
	\includegraphics[width=0.325\textwidth]{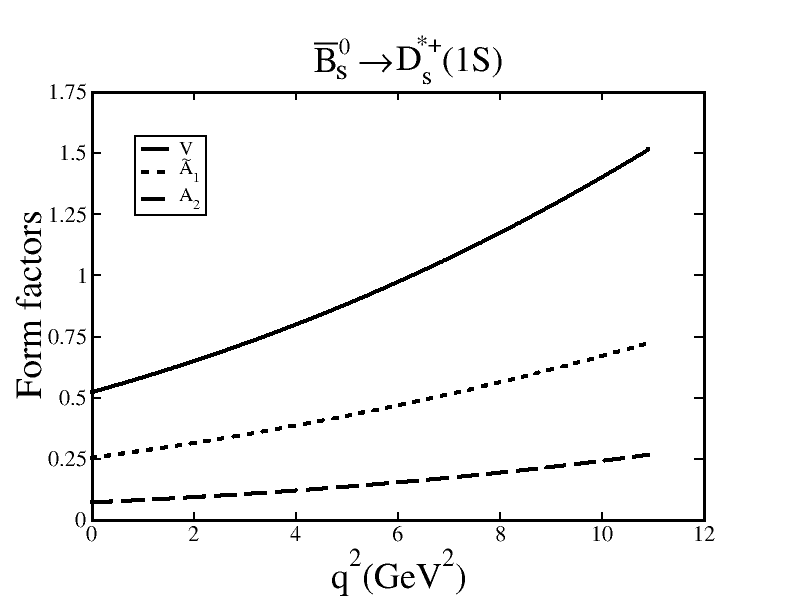}
	\includegraphics[width=0.325\textwidth]{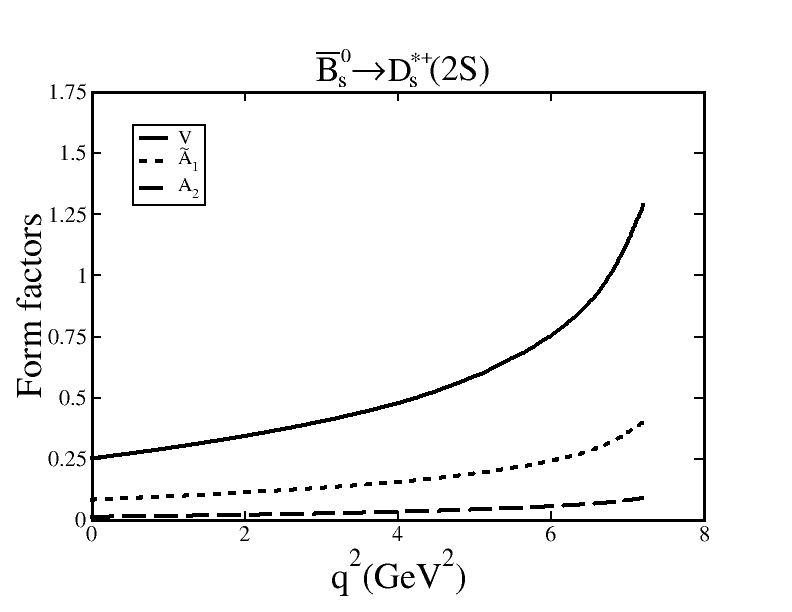}
	\includegraphics[width=0.325\textwidth]{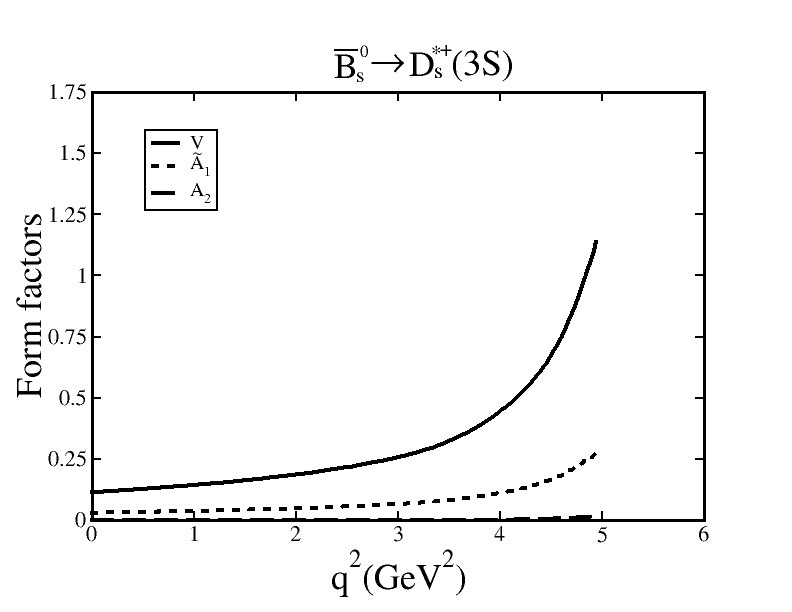}
	\caption{ $q^2$-dependence of form factors in $\bar{B_s^0}\to D_s^{*+}(nS)$ type decays.}
\end{figure}

\begin{table}[!hbt]
	\renewcommand{\arraystretch}{1}
	\centering
	\setlength\tabcolsep{7pt}
	\caption{ Predicted values of form factors for decays to 1S state.}
	\label{tab3}
	\begin{tabular}{llccc}
		
		\hline 
  Decay mode&value at&$V(q^2)$&$A_1(q^2)$&$A_2(q^2)$ \\
		
		\hline
		${B_c}^-\to J/\psi$&$q^2=0$&0.482&0.125&0.0095\\
		&$q^2=q^2_{max}$&2.121&0.452& 0.0006\\
		${B_c}^-\to D_s^{*-}$&$q^2=0$&0.077&0.016&0.0192\\
		&$q^2=q^2_{max}$&4.584&0.466&0.8853\\
		${B_c}^-\to {D}^{*-}$&$q^2=0$&0.033&0.006&0.0105\\
		&$q^2=q^2_{max}$&6.527&0.403&1.5552\\
		$B^-\to J/\psi$&$q^2=0$&0.727&0.198&0.0003\\
		&$q^2=q^2_{max}$&2.049&0.485&0.0097\\
		$B^-\to {D}^{0*}$&$q^2=0$&0.578&0.307&0.0954\\
		&$q^2=q^2_{max}$&1.447&0.605&0.2885\\
		$\bar{B^0}\to D^{*+}$&$q^2=0$&0.578&0.307&0.0953\\
		&$q^2=q^2_{max}$&1.448&0.604&0.2877\\
		$\bar{B_s^0}\to D_s^{*+}$&$q^2=0$&0.523&0.254&0.0725\\
		&$q^2=q^2_{max}$&1.514&0.584&0.2674\\
		\hline
		
	\end{tabular}
\end{table}
\begin{table}[!hbt]
	\renewcommand{\arraystretch}{1}
	\centering
	\setlength\tabcolsep{7pt}
	\caption{ Predicted values of form factors for decays to 2S state.}
	\label{tab4}
	\begin{tabular}{llccc}
		
		\hline Decay mode&value at&$V(q^2)$&$A_1(q^2)$&$A_2(q^2)$ \\
		
		\hline
		${B_c}^-\to \psi(2S)$&$q^2=0$&0.245&0.048&0.022\\
		&$q^2=q^2_{max}$&1.829&0.314&0.155\\
		${B_c}^-\to D_s^{*-}(2S)$&$q^2=0$&0.066&0.010&0.019\\
		&$q^2=q^2_{max}$&3.961&0.325&1.087\\
		${B_c}^-\to {D}^{*-}(2S)$&$q^2=0$&0.046&0.006&0.015\\
		&$q^2=q^2_{max}$&5.789&0.300&1.869\\
		$B^-\to \psi(2S)$&$q^2=0$&0.415&0.086&0.037\\
		&$q^2=q^2_{max}$&1.460&0.273&0.164\\
		$B^-\to {D}^{0*}(2S)$&$q^2=0$&0.273&0.094&0.019\\
		&$q^2=q^2_{max}$&1.231&0.357&0.097\\
		$\bar{B^0}\to D^{*+}(2S)$&$q^2=0$&0.265&0.094&0.020\\
		&$q^2=q^2_{max}$&1.224&0.363&0.106\\
		$\bar{B_s^0}\to D_s^{*+}(2S)$&$q^2=0$&0.252&0.084&0.014\\
		&$q^2=q^2_{max}$&1.287&0.360&0.090\\
		\hline
		
	\end{tabular}
\end{table}
\par With the input parameters (27,28), we first study the $q^2$-dependence of the dimensionless form factors in their allowed kinematic ranges. Our predictions do not entirely agree with HQS relation. This is due to the well-known fact that the heavy flavor symmetry is not strictly applicable in the heavy meson sector, particularly when two heavy constituent quarks are involved. The $q^2$-dependence of form factors for $B_c^-, B^-,\bar{B^0}, \bar{B_s^0}$-decays to $S$-wave ($1S, 2S, 3S$) charmonium and charm meson states in the present model is shown in the Figs. (2-8). We find that the departure from HQS relation is more and more pronounced in the decays to higher and higher excited states. This is due to different kinematics and four-momentum transfer involved in different decay modes.\\
\begin{table}[!hbt]
	\renewcommand{\arraystretch}{1}
	\centering
	\setlength\tabcolsep{7pt}
	\caption{ Predicted values of form factors for decays to 3S state.}
	\label{tab5}
	\begin{tabular}{llccc}
		
		\hline Decay mode&value at&$V(q^2)$&$A_1(q^2)$&$A_2(q^2)$ \\
		
		\hline
		${B_c}^-\to \psi(3S)$&$q^2=0$&0.1121&0.0198&0.0015\\
		&$q^2=q^2_{max}$&1.5727&0.2383&0.0015\\
		${B_c}^-\to D_s^{*-}(3S)$&$q^2=0$&0.0286&0.0036&0.0014\\
		&$q^2=q^2_{max}$&3.6723&0.2490&0.0014\\
		${B_c}^-\to {D}^{*-}(3S)$&$q^2=0$&0.0213&0.0022&0.0014\\
		&$q^2=q^2_{max}$&5.3214&0.2283&0.0014\\
		$B^-\to \psi(3S)$&$q^2=0$&0.2570& 0.0465&0.0014\\
		&$q^2=q^2_{max}$&1.2710&0.2084&0.0014\\
		$B^-\to {D}^{0*}(3S)$&$q^2=0$&0.1177&0.0337&0.0034\\
		&$q^2=q^2_{max}$&1.1429&0.2742&0.0020\\
		$\bar{B^0}\to D^{*+}(3S)$&$q^2=0$&0.1203& 0.0339&0.0029\\
		&$q^2=q^2_{max}$&1.1469& 0.2716&0.0044\\
		$\bar{B_s^0}\to D_s^{*+}(3S)$&$q^2=0$&0.1143& 0.0307&0.0009\\
		&$q^2=q^2_{max}$&1.1360&0.2569&0.0176\\
		\hline
		
	\end{tabular}
\end{table}
\begin{figure}[!hbt]
	\centering
	\includegraphics[width=.45\textwidth]{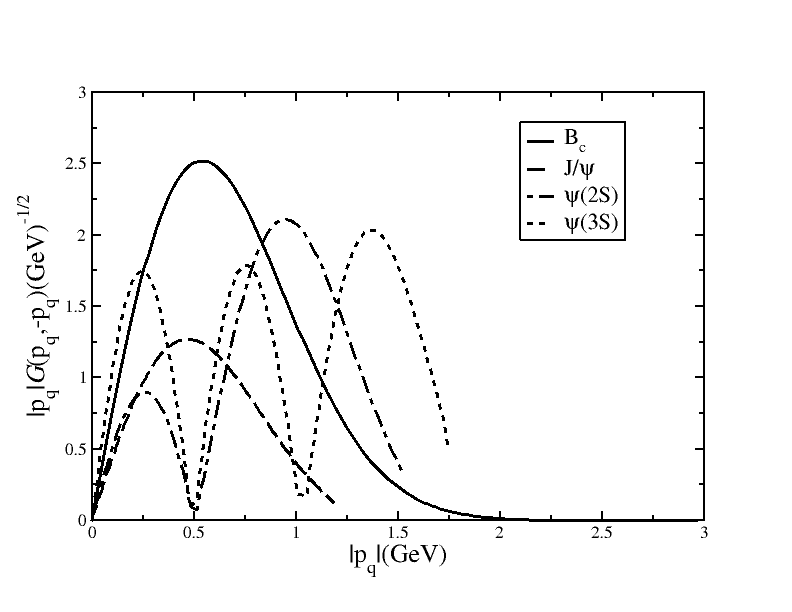}
	\includegraphics[width=.45\textwidth]{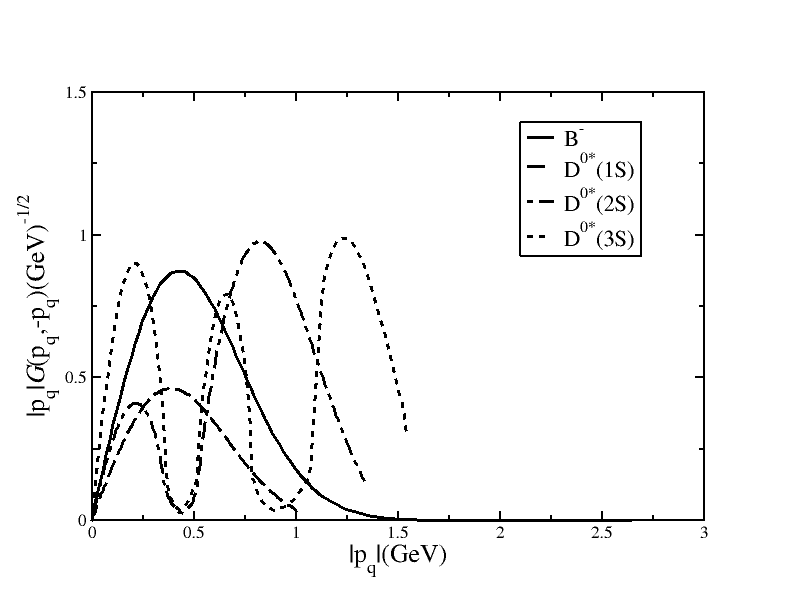}
	\caption{ Overlap of momentum distribution amplitudes of the initial and final meson states.}
\end{figure}

\begin{figure}[!hbt]
	\centering
	\includegraphics[width=.45\textwidth]{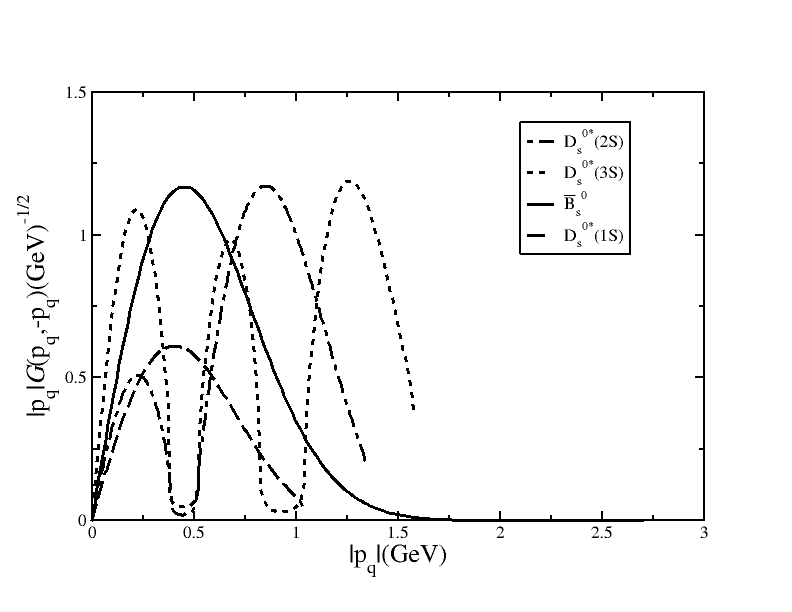}
	\includegraphics[width=.45\textwidth]{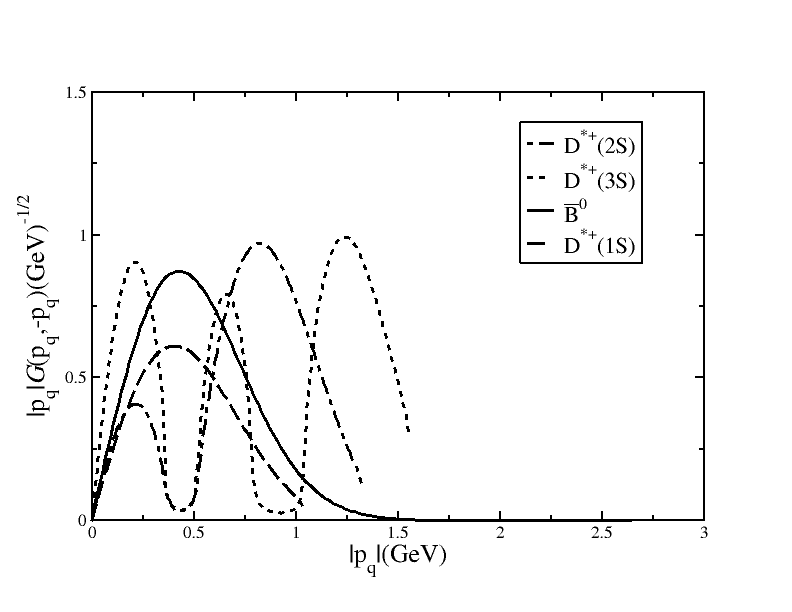}
	\caption{ Overlap of momentum distribution amplitudes of the initial and final meson states.}
\end{figure}
Our predicted form factors at $q^2\to 0$ (maximum recoil) and $q^2\to q^2_{max}$ (minimum recoil) point for the transition to $1S$, $2S$ and $3S$ charmonium and charm meson states are shown, respectively in Table (3,4,5). Before predicting the physical quantities of interest: decay width and branching fraction etc., it is interesting to go for a qualitative assessment of transition probabilities for transition to different $S$-wave states. For this, we study the radial quark momentum distribution amplitude $|\vec{p}_q|{\cal G}_{B_F}(\vec{p}_q,-\vec{p}_q)$ of the parent and daughter mesons over the physical range of respective quark momentum $\vec{p}_q$, for each such decay mode. From the plot shown in Figs. (9-10) the overlap region between the momentum distribution profile of the parent and daughter meson in the transition to the $1S$ state is found to be maximum and it decreases for transitions to higher excited $2S$ and $3S$ states. Since the invariant form factors are evaluated from overlapping integrals of participating meson wave function, it is expected that the contribution of the form factors to decay width/ branching fractions should be in decreasing order of magnitude, going from decays to $1S$ state to higher excited ($2S$ and $3S$) states.

\begin{table*}[hbt!]
	\renewcommand{\arraystretch}{1}
	\centering
	\setlength\tabcolsep{0.5pt}
	\caption{Decay widths in units of $10^{-15}$ GeV in terms of Wilson coefficients $a_1$ and $a_2$. }
	\label{tab6}
	\resizebox{\textwidth}{!}
	{
		\begin{tabular}{c|c|c|c|c|c}
			
			\hline
			$B_c^-\to J/\psi X^-$&Width&$B_c^-\to\psi(2S) X^-$&Width&$B_c^-\to\psi(3S) X^-$&Width\\		
			\hline
			$\rho^-$&1.8565$a_1^2$&$\rho^-$&0.2076$a_1^2$&$\rho^-$&0.0266$a_1^2$\\		
			$K^{*-}$&0.1123$a_1^2$&$K^{*-}$&0.0127$a_1^2$&$K^{*-}$&0.0016$a_1^2$\\
			$D_s^{*-}$&$(3.124a_1+3.529a_2)^2$&$D_s^{*-}$&$(1.344a_1+3.6702a_2)^2$&$D_s^{*-}$&$(0.942a_1+3.502a_2)^2$\\
			$D^{*-}$&$(0.758a_1+0.359a_2)^2$&$D^{*-}$&$(0.331a_1+0.842a_2)^2$&$D^{*-}$&$(0.191a_1+0.847a_2)^2$\\
			\hline
			$\bar{B_s^0}\to D_s^{*+}(1S)X^-$&Width&$\bar{B_s^0}\to D_s^{*+}(2S)X^-$&Width&$\bar{B_s^0}\to D_s^{*+}(3S)X^-$&Width\\
			\hline
			$\rho^-$&3.929$a_1^2$&$\rho^-$&0.3502$a_1^2$&$\rho^-$&0.0350$a_1^2$\\	
			$K^{*-}$&\ 0.2319$a_1^2$&$K^{*-}$&0.0211$a_1^2$&$K^{*-}$&0.0021$a_1^2$\\
			\hline
			$\bar{B^0}\to D^{*+}(1S)X^-$&Width&$\bar{B^0}\to D^{*+}(2S)X^-$&Width&$\bar{B^0}\to D^{*+}(3S)X^-$&Width\\
			\hline
			$\rho^-$&4.918$a_1^2$&$\rho^-$&0.386$a_1^2$&$\rho^-$&0.0367$a_1^2$\\	
			$K^{*-}$&\ 0.2885$a_1^2$&$K^{*-}$&\ 0.0232$a_1^2$&$K^{*-}$&0.0022$a_1^2$\\
			\hline
			$B^-\to D^{*0}(1S)X^-$&Width&$B^-\to D^{*0}(2S)X^-$&Width&$B^-\to D^{*0}(3S)X^-$&Width\\
			\hline	
			$\rho^-$&4.931$a_1^2$&$\rho^-$&0.3787$a_1^2$&$\rho^-$&0.0371$a_1^2$\\	
			$K^{*-}$&0.289$a_1^2$&$K^{*-}$&0.0228$a_1^2$&$K^{*-}$&0.0023$a_1^2$\\
			\hline
			$B^-\to J/\psi X^-$&Width&$B^-\to \psi(2S)X^-$&Width&$B^-\to \psi(3S)X^-$&Width\\
			\hline
			$\rho^-$&1.032$a_2^2$&$\rho^-$&0.892$a_2^2$&$\rho^-$&\ 0.628$a_2^2$\\	
			$K^{*-}$&31.29$a_2^2$&$K^{*-}$&\ 20.912$a_2^2$&$K^{*-}$&12.22$a_2^2$\\
			\hline
			
	\end{tabular}
	}	
\end{table*}

\begin{table*}[!hbt]
	\renewcommand{\arraystretch}{1}
	\centering
	\setlength\tabcolsep{0.5pt}
	\caption{ Branching fractions in $\%$ for values of Wilson coefficients $a_1$ and $a_2$. }
	\label{tab7}
	\resizebox{\textwidth}{!}
	{
	\begin{tabular}{cccccclclll}
		
		\hline &\multicolumn{3}{c}{Branching Fraction}&&\cite{A20}&\cite{A23}&\cite{A25}&\cite{A49}&\ \cite{A89}&\\
		\cline{2-4}
		$B_c^-\to J/\psi X^-$ &$a_1$=1.12&$a_1$=1.14&$a_1$=0.93&&&&&&&\\
		& $a_2$=-0.26&$a_2$=-0.20&$a_2$=-0.27&&&&&&&\\
		\hline 
		$ \rho^-$&0.1804&0.186&0.124&&\ 0.16&\ 0.49&&\  0.40&&\\
		$K^{*-}$&0.0109&0.011&\ 0.0075&&\  0.01&\ 0.028&\ \ \ $0.0109\pm 0.0033$&\  0.022&&\\
		$ D_s^{*-}$&0.5159&\ 0.6317&\ 0.2953&&&\ 0.97&$0.412\pm 0.123$&\ 0.67&\ 0.78&\\
		$ D^{*-}$&0.0442&\ 0.0486&\ 0.0286&&&\ 0.045&\ \ \ $0.0182\pm 0.0055$&\ 0.028&\ 0.031&\\ 
		\hline
		$\bar{B_s^0}\to D_s^{*+}(1S)X^-$&&Branching Fraction&&\cite{A18}&&&&&&\cite{A82}\\
		\hline
		$ \rho^-$&1.1734&1.2157&\ 0.8091&$0.726^{+0.076}_{-0.071}$&&&&&&$0.96\pm0.21$\\
		$ K^{*-}$&0.0692&0.0717&\ 0.0477&\ \ \ $0.0688^{+0.0067}_{-0.0064}$&&&&&&\\ 
		\hline 
	    $\bar{B^0}\to D^{*+}(1S)X^-$&&Branching Fraction&&\cite{A18}&&&&&&
     \cite{A82}\\
		\hline
		$\rho^-$&1.4238&1.4751&\ 0.9817&$0.873^{+0.078}_{-0.073}$&&&&&&$0.68\pm0.09$\\
		$K^{*-}$&0.0834&0.0864&\ 0.0575&\ \ \ $0.0758^{+0.0064}_{-0.0059}$&&&&&&$0.033\pm0.006$\\  
		\hline
		$B^-\to D^{*0}(1S)X^-$&&Branching Fraction&&\cite{A18}&&&&&&\cite{A82}\\
		\hline                     
		$\rho^-$&1.5391&1.5946&\ 1.0612&$0.873^{+0.079}_{-0.073}$&&&&&&$0.98\pm0.17$\\
		$K^{*-}$&0.0901&0.0934&\ 0.0621&\ \ \ $0.0846^{+0.0070}_{-0.0064}$&&&&&&$0.081\pm0.014$\\
		\hline
		$B^-\to J/\psi X^-$&&Branching Fraction&&&&&&&&\cite{A82}\\
		\hline
		$\rho^-$&0.0173&0.0102&0.0187&&&&&&&$0.0041\pm 0.0005$\\
		$K^{*-}$&0.5264&0.3115&0.5677&&&&&&&$0.143\pm 0.008$\\
		\hline
		
	\end{tabular}
}
\end{table*}
\par We calculate the decay width from the expression (26) via (14) and (9-10) and our predicted decay widths $\Gamma (B_F\to V_1(nS)V_2)$ for general values of QCD coefficients $(a_1, a_2)$ of the operator product expansion are listed in Table 6 to facilitate a comparison with other dynamical model predictions. Our predicted branching fractions (BFs) for  $B_c^-, \bar{B_s^0}, B^-, \bar{B^0}$-decays to $1S$, $2S$ and $3S$ charmonium and charm meson states, are listed in Table (7,8,9), respectively, in reasonable agreement with available experimental data \cite{A82} and other model predictions.
\noindent Our results for BFs of decays to $1S$, $2S$ and $3S$ states corresponding to 3 sets of QCD parameters are listed in the second column of each table. As expected, our predicted branching fractions are obtained in the hierarchy:\\
\begin{eqnarray}
   {\cal B}(B_F\to V_1(3S)V_2) &&<{\cal B}(B_F\to V_1(2S)V_2)\nonumber \\&&<{\cal B}(B_F\to V_1(1S)V_2)\nonumber. 
\end{eqnarray}
\begin{table*}[!hbt]
	\renewcommand{\arraystretch}{0.8}
	\centering
	\setlength\tabcolsep{0.35pt}
	\caption{ Branching fractions in $10^{-4}$ for values of Wilson coefficients $a_1$ and $a_2$. }
	\label{tab8}
	\resizebox{\textwidth}{!}
	{
	\begin{tabular}{ccccccccc} 
		\hline&\multicolumn{3}{c}{Branching Fraction}&\cite{A15}&\cite{A18}&\ \cite{A20}&\cite{A25}&\\
		\cline{2-4}
		$B_c^-\to\psi(2S) X^-$ &$a_1$=1.12&$a_1$=1.14&$a_1$=0.93&&&& &\\
		& $a_2$=-0.26&$a_2$=-0.20&$a_2$=-0.27&&&&&\\
		\hline 
		$ \rho^-$&2.01&2.09&1.39&5.69&$4.04^{+0.21}_{-0.17}$&\ 1.8&&\\
		$K^{*-}$&\ 0.124&\ 0.128&\ 0.085&\ 0.304&\ \ \ \ \ $0.2822^{+0.0156}_{-0.0126}$&\  1.0&\ \ \ $0.439\pm0.071$&\\
		$ D_s^{*-}$&2.35&4.92&0.52&&&&$8.85\pm 2.54$&\\
		$ D^{*-}$&\ 0.178&0.33&0.05&&&&\ \ \ $0.432\pm 0.117$&\\ 
		\hline
		$\bar{B_s^0}\to D_s^{*+}(2S)X^-$&&Branching Fraction&&&\cite{A18}&&\ \cite{A26}&\\
		\hline
		$ \rho^-$&\ 10.45&\ 10.83&7.21&&\ $0.475^{+0.164}_{-0.121}$&&\ 22&\\
		$ K^{*-}$&0.63&0.65&0.43&&\ \ \ \  $0.0332^{+0.0122}_{-0.0090}$&&\ \  1.2&\\ 
		\hline 
		$\bar{B^0}\to D^{*+}(2S)X^-$&&Branching Fraction&&&\cite{A18}&&&\\
		\hline
		$\rho^-$&\ 11.17&\ 11.57&\ 7.706&&\ $0.267^{+0.081}_{-0.061}$&&&\\
		$K^{*-}$&0.67&0.69&0.46&&\ \ \ \ $0.0162^{+0.0050}_{-0.0038}$&&&\\  
		\hline
		$B^-\to D^{*0}(2S)X^-$&&Branching Fraction&&&\cite{A18}&&&\\
		\hline                     
		$\rho^-$&\ 11.81&\ 12.24&8.14&&\ $0.287^{+0.088}_{-0.066}$&&&\\
		$K^{*-}$&0.70&0.72&0.48&&\ \ \ \ $0.0173^{+0.0054}_{-0.0041}$&&&\\
		\hline
		$B^-\to \psi(2S) X^-$&&Branching Fraction&&&&&\cite{A82}&\\
		\hline
		$\rho^-$&1.49&0.88&1.61&&&&&\\
		$K^{*-}$&\ 35.18&\ 20.81&\ 37.93&&&&$6.7\pm1.4$&\\
		\hline
			\end{tabular}
		}
\end{table*}

\begin{table}[!hbt]
	\renewcommand{\arraystretch}{1}
	\centering
	\setlength\tabcolsep{0.5pt}
	\caption{ Branching fractions in $10^{-5}$ for values of Wilson coefficients $a_1$ and $a_2$. }
	\label{tab9}
	\resizebox{0.5\textwidth}{!}
	{
	\begin{tabular}{ccccc}
		 
		\hline&\multicolumn{3}{c}{Branching Fraction}&\cite{A18}\\
		\cline{2-4}
		$B_c^-\to\psi(3S) X^-$ &$a_1$=1.12&$a_1$=1.14&$a_1$=0.93& \\
		& $a_2$=-0.26&$a_2$=-0.20&$a_2$=-0.27&\\
		\hline 
		$ \rho^-$&2.5&2.6&1.7&$3.35^{+0.78}_{-0.68}$\\
		$K^{*-}$&\ \ 0.16&\  0.17&\ 0.11&\ \ $0.229^{+0.055}_{-0.045}$\\
		$ D_s^{*-}$&1.6& 10.8&\ 0.37&\\
		$ D^{*-}$&\ \ \ 0.003&\ \ 0.181&\ \ 0.202&\\ 
		\hline
		$\bar{B_s^0}\to D_s^{*+}(3S)X^-$&&Branching Fraction&&\cite{A18}\\
		\hline
		$ \rho^-$&10.4&\ 10.85&7.2&$3.55^{+0.88}_{-0.69}$\\
		$ K^{*-}$&0.65&0.67&\ 0.44&\ \ $0.251^{+0.061}_{-0.048}$\\ 
		\hline 
		$\bar{B^0}\to D^{*+}(3S)X^-$&&Branching Fraction&&\cite{A18}\\
		\hline
		$\rho^-$&10.6&\ 11.03&7.34&$1.03^{+0.44}_{-1.04}$\\
		$K^{*-}$&0.66&0.68&0.45&\ \ \ \ \ $0.0709^{+0.0310}_{-0.0719}$\\  
		\hline
		$B^-\to D^{*0}(3S)X^-$&&Branching Fraction&&\cite{A18}\\
		\hline                     
		$\rho^-$&11.5&12&7.9&$1.11^{+0.50}_{-1.12}$\\
		$K^{*-}$&0.72&\ 0.74&\ 0.49&\ \ \ \ \  $0.0753^{+0.0347}_{-0.0766}$\\
		\hline
		$B^-\to \psi(3S) X^-$&&Branching Fraction&&\\
		\hline
		$\rho^-$&10.5&6.2&11.3&\\
		$K^{*-}$&\ \  205.5&\ \  121.6&\ \ 221.6&\\
		\hline
		
		\end{tabular}
  }
	\end{table}
\noindent Our results for transitions to $2S$ and $3S$ states are obtained two and three orders of magnitude down compared to those obtained for transition to $1S$ state. The node structure of the $2S$ wave function is responsible for small BFs. Since there is no node for the initial wave function, the contribution from the positive and negative parts of the final wave function cancel each other out yielding small BFs. In the case of the transition to $3S$ states, there are even more serious cancellations; leading to still smaller BFs. As expected, the tighter phase space and the $q^2$-dependence of the form factors typical to the decay mode lead to smaller BFs for transitions to higher excited $2S$ and still smaller for the transition of $3S$ states.
\par The BFs of the decay modes, considered in the present study, are obtained in a wide range of $\sim 10^{-2}-10^{-6}$. For nonleptonic $B_F$-meson decays to 1S, 2S, and 3S charmonium and charm meson states, BFs range from $\sim 10^{-2}-10^{-3}$, $\sim 10^{-4}-10^{-5}$ and $\sim 10^{-5}-10^{-6}$, respectively. The dominant decay modes: $B_c^-\to J/\psi D_s^{*-}$, $\bar{B_s^0}\to D_s^{*+}\rho^-$, $\bar{B^0}\to D^{*+}\rho^-$, $B^-\to D^{*0}\rho^-$ and $B^-\to J/\psi K^{*-}$ to $1S$ charmonium and charm meson states are found to have BFs of 0.52, 1.17, 1.42, 1.54 and 0.53, respectively in the order ${\cal O}(\sim 10^{-2})$ which should be experimentally accessible. For such decays to corresponding $2S$ modes, the predicted BFs, of 2.35, 10.45, 11.17, 11.81 and 35.18 in the order of magnitude ${\cal O}(\sim 10^{-4})$, lie within the detection accuracy of current experiments. The neutral $B$-meson decay in the present study is found to have smaller BFs than those of charged $B$-meson decays, as expected. This may be due to the spectator interaction effects of $d$ and $u$ quarks.
\begin{table}[!hbt]
	\renewcommand{\arraystretch}{1.3}
	\centering
	\setlength\tabcolsep{0.6pt}
	\caption{Decay widths in units of $10^{-15}$ GeV and branching fractions in $\%$ for values of Wilson coefficient $a_1$ and $a_2$.}
	\label{tab10}
 \resizebox{0.5\textwidth}{!}
	{
	\begin{tabular}{lclllc}		
	\hline &&\multicolumn{3}{c}{Branching Fraction}&\cite{A82}\\	
\cline{3-5}	
		Decay modes &Decay width&\ $a_1$=1.12&\ $a_1$=1.14&\ $a_1$=0.93 &\\
		&&\ $a_2$=-0.26&\ $a_2$=-0.20&\ $a_2$=-0.27&\\
		\hline 
		$\bar{B^0_s}\to D_s^{*+}(1S)D^{*-}$&0.8116$a_1^2$&\ 0.2423&\ 0.2510&\ 0.1670&\\
		$\bar{B^0_s}\to D_s^{*+}(2S)D^{*-}$&0.1052$a_1^2$&\ 0.0313&\ 0.0324&\ 0.0216&\\
		$\bar{B^0_s}\to D_s^{*+}(3S)D^{*-}$&0.0304$a_1^2$&\ 0.0090&\ 0.0094&\ 0.0062&\\
		\hline
		$\bar{B^0}\to D^{*+}(1S)D^{*-}$&1.1406$a_1^2$&\ 0.2631&\ 0.2726&\ 0.1814&\\
		$\bar{B^0}\to D^{*+}(2S)D^{*-}$&0.1077$a_1^2$&\ 0.0311&\ 0.0323&\ 0.0215&\\
		$\bar{B^0}\to D^{*+}(3S)D^{*-}$&0.0329$a_1^2$&\ 0.0095&\ 0.0098&\ 0.0065&\\
		\hline
		$\bar{B^0}\to D^{*+}(1S)D_s^{*-}$&21.924$a_1^2$&\ 6.3460&\ 6.5740&\ 4.3750&$0.8\pm 0.11$\\
		$\bar{B^0}\to D^{*+}(2S)D_s^{*-}$&2.7331$a_1^2$&\ 0.7911&\ 0.8196&\ 0.5454&\\
		$\bar{B^0}\to D^{*+}(3S)D_s^{*-}$&1.1440$a_1^2$&\ 0.3312&\ 0.3432&\ 0.2284&\\
		\hline
		$B^-\to D^{*0}(1S)D^{*-}$&0.9118$a_1^2$&\ 0.2846&\ 0.2948&\ 0.1962&\ \ \ $0.081\pm 0.017$\\
		$B^-\to D^{*0}(2S)D^{*-}$&0.1072$a_1^2$&\ 0.0334&\ 0.0346&\ 0.0230&\\
		$B^-\to D^{*0}(3S)D^{*-}$&0.0305$a_1^2$&\ 0.0095&\ 0.0098&\ 0.0065&\\
		\hline
		$B^-\to D^{*0}(1S)D_s^{*-}$&21.984$a_1^2$&\ 6.8625&\ 7.1098&\ 4.7316&$1.71\pm 0.24$\\
		$B^-\to D^{*0}(2S)D_s^{*-}$&2.7074$a_1^2$&\ 0.8451&\ 0.8756&\ 0.5827&\\
		$B^-\to D^{*0}(3S)D_s^{*-}$&1.0567$a_1^2$&\ 0.3298&\ 0.3417&\ 0.2274&\\
		\hline
		
	\end{tabular}
 }
\end{table}
\par As discussed earlier, the nonleptonic $B_F$-decays to two charmful vector meson ($V_1V_2$) states are of special interest as they help to evaluate $CP$-asymmetry factors, which provide the clue for testing SM predictions and exploring possible new physics beyond SM. The predicted decay widths (in $10^{-15} $ GeV) for general values of QCD parameters ($a_1,a_2$) for $B^-$, $\bar{B^0}$ and $\bar{B_s^0}$ decays to two charmful ($1S, 2S, 3S$) states along with their BFs, obtained in order of magnitude $\sim 10^{-2}-10^{-4}$, are shown in Table 10. However, the predicted  BFs for $B_c^-$-decays: $B_c^-\to D^{*-}_{(s)}D^{*0}$ and $B_c^-\to \bar{D}^{*0}D^{*-}_{(s)}$, obtained in the order of magnitude ${\cal O}(10^{-6})$, are shown in Table 11 and 12, which cannot be measured in current experiments. Our predicted BFs for $\bar{B^0}$, $B^-$, $B_c^-$ decays to two charmful states, however, are found somewhat underestimated compared to predictions to Ref. \cite{A23} and available experimental data \cite{A82}.\\
\noindent The relative size of BFs for nonleptonic decays is broadly estimated from a power counting of QCD factors: ($a_1$, $a_2$) in the Wolfenstein parameterization \cite{A90}. Accordingly, class I decays determined by $a_1$ are found to have comparatively large BFs as shown in Table. (7,8,9,10). On the other hand, class II decay modes, determined by $a_2$, are found to have relatively small BFs Table. (7,8,9,11), as expected, except  for decay modes: $B^-\to J/\psi K^{*-}$ and $B^-\to \psi(2S) K^{*-}$, characterized by a product of CKM factors: $V_{bc}V_{cs}$, which have BFs $\sim 0.31\%$ and $0.12\%$, respectively. These modes should be measured at high luminosity hadron colliders. In class III decay modes that are characterized by Pauli interference, the BFs are determined by the relative value of $a_1$ with respect to $a_2$. Considering positive values of $a_1=1.12$ and negative value of $a_2=-0.26$ in Set 1, for example, which leads to destructive interference, the decay modes are suppressed compared to the case where interference is switched off. However, at a  qualitative level, where the ratio $\frac{a_2}{a_1}$, a function of running coupling constant $\alpha_S$ evaluated at the factorization scale, is shown to be positive in the case of $b$-flavored meson decays corresponding to small coupling \cite{A32}. The experimental data also favor constructive interference of the color-favored and color-suppressed $b$-flavored meson decay modes. Considering positive value of $a_2^b=0.26$, our predicted BFs of class III decay modes: $B_c^-\to J/\psi D_s^{*-}$ and $B_c^-\to \psi(2S) D_s^{*-}$ find enhancement by a factor of $\sim 3$ and $\sim 20$, respectively, over that obtained with $a_2=-0.26$. For decay modes to $3S$ states, the enhancement is still more significant.
\begin{table}[!hbt]
	\renewcommand{\arraystretch}{1.3}
	\centering
	\setlength\tabcolsep{0.5pt}
	\caption{Decay widths in units of $10^{-15}$ GeV and branching fractions in $10^{-6}$ for values of Wilson coefficient $a_1$ and $a_2$.}
	\label{tab11}
 \resizebox{0.5\textwidth}{!}
	{
	\begin{tabular}{cllllc}
		\hline &&\multicolumn{3}{c}{Branching Fraction}&\ \ \ \cite{A23}\\
		
		\cline{3-5}
		Decay modes&Decay width&\ $a_1$=1.12&\ $a_1$=1.14&\ $a_1$=0.93& \\
		&&\ $a_2$=-0.26&\ $a_2$=-0.20&\ $a_2$=-0.27&\\
		\hline 
		${B^-_c}\to D_s^{*-}(1S)D^{*0}$&\ 9.6099${a_2^2}$&\ 0.5033&\ 0.2978&\ 0.5428&\ \ \ 1.6\\
		${B^-_c}\to D_s^{*-}(2S)D^{*0}$&\ 2.2700${a_2^2}$&\ 0.1189&\ 0.0703&\ 0.1282&\\
		${B^-_c}\to D_s^{*-}(3S)D^{*0}$&\ 0.3628${a_2^2}$&\ 0.0190&\ 0.0112&\ 0.0204&\\
		\hline
		${B^-_c}\to D^{*-}(1S)D^{*0}$&\ 0.1440${a_2^2}$&\ 0.0075&\ 0.0044&\ 0.0081&\ \ \ 21\\
		${B^-_c}\to D^{*-}(2S)D^{*0}$&\ 0.0664${a_2^2}$&\ 0.0034&\ 0.0020&\ 0.0037&\\
		${B^-_c}\to D^{*-}(3S)D^{*0}$&\ 0.0106${a_2^2}$&\ 0.0005&\ 0.0003&\ 0.0006&\\
		\hline
	\end{tabular}
 }
\end{table}
\begin{table}[!hbt]
	\renewcommand{\arraystretch}{1.4}
	\centering
	\setlength\tabcolsep{0.5pt}
	\caption{ Decay widths in units of $10^{-15}$ GeV and branching fractions of order $10^{-6}$ for general values of the Wilson coefficients $a_1$ and  $a_2$.}
	\label{tab12}
	\resizebox{0.5\textwidth}{!}
	{
		\begin{tabular}{lllll}
		
			\hline&&\multicolumn{3}{c}{Branching Fraction}\\
			\cline{3-5}
			\ \ \ Decay mode &\ \ \ \ Decay width \ &\  $a_1$=1.12&\ $a_1$=1.14&\ $a_1$=0.93 \\
			&&\ $a_2$=-0.26&\ $a_2$=-0.20&\ $a_2$=-0.27\\
			\hline
			${B_c}^-\to\bar{D}^{*0}(1S)D_s^{*-}$&$(0.0648a_1+0.0979a_2)^2$&\ 1.72&\ 2.28&\ 0.887\\
			${B_c}^-\to\bar{D}^{*0}(2S)D_s^{*-}$&$(0.0438a_1+0.1450a_2)^2$&\ 0.1007&\ 0.3416 &\ 0.002\\
			${B_c}^-\to\bar{D}^{*0}(1S)D^{*-}$&$(0.0120a_1+0.0119a_2)^2$&\ 0.082&\ 0.098&\ 0.0489\\
			${B_c}^-\to\bar{D}^{*0}(2S)D^{*-}$&$(0.0049a_1+0.0212a_2)^2$&\ 0.4463$\times10^{-6}$&\ 0.0014&\ 0.0010\\
			\hline
		
		\end{tabular}
	}
\end{table}
\par In the spirit of the experimental data favoring a constructive interference of the  color-favored and color-suppressed $b$-flavored meson decays, the effect of Pauli interference inducing enhancement of BFs can be further probed by casting the decay width ($\Gamma$) in the form: $\Gamma=\Gamma_0+\Delta\Gamma$, where $\Gamma_0=x_1^2a_1^2+x_2^2a_2^2$, $\Delta\Gamma=2x_1x_2a_1a_2$ and then evaluating $\frac{\Delta\Gamma}{\Gamma_0}$ in each case as done in \cite{A33,A34,A35,A36,A37,A38,A39,A40,A55}. We find that the absolute values of $\frac{\Delta\Gamma}{\Gamma_0}$ for $B_c^-\to J/\psi D_s^{*-}, B_c^-\to\psi(2S) D_s^{*-})$,  $B_c^-\to J/\psi D^{*-}$ and $B_c^-\to\psi(2S) D^{*-}$ (in $\%$) are $\sim 49$, $\sim 90$, $\sim 22$, and $\sim 88$, respectively. For $B_c$-decays to two charmful states: $B_c^-\to \bar{D}^{*0} D_s^{*-}$, $B_c^-\to\bar{D}^{*0}(2S) D_s^{*-}$, $B_c^-\to \bar{D}^{*0} D^{*-}$, $B_c^-\to\bar{D}^{*0}(2S) D^{*-})$, the enhancement in $\%$ is found to be $\sim 63$, $\sim 96$, $\sim 44$, and $\sim 100$, respectively. This indicates that interference is more significant in $B_c^-$-decays to two charmful states: $B_c^-\to\bar{D}^{*0}(1S, 2S)D^{*-}_{(s)}$ compared to other decay modes. This is particularly important since such decay modes have been proposed \cite{A91,A92,A93,A94} for extraction of the CKM angle $\gamma$ through amplitude relation.
\begin{table}[!hbt]
	\renewcommand{\arraystretch}{1.3}
	\centering
	\setlength\tabcolsep{5pt}
	\caption{Predicted longitudinal fraction ($R_L$) and $CP$-odd fraction ($R_\perp$). }
	\label{tab13}
	\begin{tabular}{lcc}
		
		\hline
		&Longitudinal&\\
		&polarization&\ $CP$ -odd\\
		Decay modes &   fraction($R_L$)& \ fraction($R_\perp$)\\
		\hline
		${B^-_c}\to D^{*0}(1S)D_s^{*-}$&0.796&0.133\\
		${B^-_c}\to D^{*0}(2S)D_s^{*-}$&0.688&0.189\\
		${B^-_c}\to D^{*0}(3S)D_s^{*-}$&0.603&  0.202\\
		\hline
		${B^-_c}\to D^{*0}(1S)D^{*-}$&0.797&0.156\\
		${B^-_c}\to D^{*0}(2S)D^{*-}$&0.690&0.224\\
		${B^-_c}\to D^{*0}(3S)D^{*-}$&0.606&0.248\\
		\hline	
		${B^-_c}\to \bar{D}^{*0}(1S)D_s^{*-}$&0.786&0.148\\
		${B^-_c}\to \bar{D}^{*0}(2S)D_s^{*-}$&0.640&0.234\\
		\hline
		${B^-_c}\to \bar{D}^{*0}(1S)D^{*-}$&0.797&0.156\\
		${B^-_c}\to \bar{D}^{*0}(2S)D^{*-}$&0.638&0.276\\
		\hline
	\end{tabular}	
\end{table}
\begin{table}[!hbt]
	\renewcommand{\arraystretch}{1.3}
	\centering
	\setlength\tabcolsep{5pt}
	\caption{Predicted longitudinal fraction ($R_L$) and $CP$-odd fraction ($R_\perp$) of $\bar{B^0_s}$,$\bar{B^0}$,$B^-$. }
	\label{tab14}
	\begin{tabular}{lcc}
		\hline
		&Longitudinal&\\
		&polarization&\ $CP$ -odd\\
		Decay modes &   fraction($R_L$)& \ fraction($R_\perp$)\\
		\hline $\bar{B^0_s}\to D_s^{*+}(1S)D^{*-}$&0.677&0.0627\\
		$\bar{B^0_s}\to D_s^{*+}(2S)D^{*-}$&0.524&0.0597\\
		$\bar{B^0_s}\to D_s^{*+}(3S)D^{*-}$&0.396&0.0291\\
		\hline
		$\bar{B^0}\to D^{*+}(1S)D^{*-}$&0.675&0.0606\\
		$\bar{B^0}\to D^{*+}(2S)D^{*-}$&0.514&0.0571\\
		$\bar{B^0}\to D^{*+}(3S)D^{*-}$&0.381&0.0230\\
		\hline
		$\bar{B^0}\to D^{*+}(1S)D_s^{*-}$&0.646&0.0629\\
		$\bar{B^0}\to D^{*+}(2S)D_s^{*-}$&0.482&0.0537\\
		$\bar{B^0}\to D^{*+}(3S)D_s^{*-}$&0.350&0.0094\\
		\hline
		$B^-\to D^{*0}(1S)D^{*-}$&0.675&0.0605\\
		$B^-\to D^{*0}(2S)D^{*-}$&0.502&0.0554\\
		$B^-\to D^{*0}(3S)D^{*-}$&0.390&0.0263\\
		\hline
		$B^-\to D^{*0}(1S)D_s^{*-}$&0.647&0.0629\\
		$B^-\to D^{*0}(2S)D_s^{*-}$&0.470&0.0514\\
		$B^-\to D^{*0}(3S)D_s^{*-}$&0.358&0.0137\\
		\hline
	
	\end{tabular}	
\end{table}
\par Another area of interest is $CP$-violation in nonleptonic decay of $b$-flavored mesons to two charmful states. The evaluation of the $CP$-odd fraction $R_\perp$ indeed indicates the degree of $CP$-violation in a decay process. We predict the longitudinal polarization fraction  $R_L$ and $CP$-odd fraction $R_\perp$ from their expressions in terms of positive, negative and longitudinal polarization (10) as:
\begin{eqnarray}
	R_L&=&\frac{|{\cal A}_{ll}|^2}{[|{\cal A}_\pm|^2+|{\cal A}_\mp|^2+|{\cal A}_{ll}|^2]},\nonumber\\
	R_\perp&=&\frac{|{\cal A}_\pm-{\cal A}_\mp|^2}{2[|{\cal A}_\pm|^2+|{\cal A}_\mp|^2+|{\cal A}_{ll}|^2]}.
\end{eqnarray}
Our predicted $R_L$ and $R_\perp$ for $(B_c^-\to D^{*0}(nS)D^{*-}_{(s)})$ and ($\bar{B^0}\to D^{*+}(nS)D^{*-}_{(s)}$, $B^-\to D^{*0}(nS)D^{*-}_{(s)}$, $\bar{B_s^0}\to D^{*+}_{s}(nS) D^{*-}$) are shown in Table 13 and Table 14, respectively. In all decay modes to two charmful states considered here, the longitudinal polarization fractions$(R_L)$ dominate over the transverse polarization fraction $(R_\perp)$. However the $CP$-odd fractions $(R_\perp)$ in nonleptonic $B_c$-decays to two charmful states are obtained here one order of magnitude higher than that in such other $b$-flavored meson decays; which indicates that $CP$-violation is more pronounced in $B_c^-\to D^{*0}D^{*-}_{(s)}$ decays compared to that obtained in corresponding decays of ($\bar{B^0}$, $\bar{B_s^0}$ and $B^-$) mesons. It is also found that $R_\perp$ in $B_c^-$-sector gets enhanced going from decays to ground state to corresponding decays to higher excited states ($2S$ and $3S$). Apart from significant $CP$-odd fraction $R_\perp$ predicted in $B_c^-$-decays to two charmful states, its noticeable enhancement is also predicted for $B_c^-$ and $B^-$ decays to charmonium states: $B_c^-\to \psi(nS)D^{*-}$, $B_c^-\to \psi(nS)D_s^{*-}$, $B^-\to \psi(nS)\rho^-$ and $B^-\to \psi(nS)K^{*-}$ as shown in Table 15.
\par  For color-favored $B_c^-\to D^*(D^{*-}, D_s^{*-})$ decays, the effect arising due to the short-distance non-spectator contribution is shown to be marginal \cite{A95}. However, the long-distance (LD) nonfactorizable contributions from rescattering effects, final-state interactions, etc., may not be negligible. If a significant LD effect exists, one expects a large $CP$-odd fraction in these decays. The predicted longitudinal and transverse helicity amplitudes and the form factor $g(q^2)$ yield $R_\perp$ values for different $B_F\to  V_1(nS)V_2$ decays, which are shown in Table 15. In particular, the predicted $R_\perp$ values for the transitions with two charmful final states indicate nonvanishing LD contributions, which lead to CP-violation in $B_c^-\to D^{*0}D^{*-}_{(s)}$ and $\bar{D}^{*0}D^{*-}_{(s)}$ decays.
\begin{table}[!hbt]
	\renewcommand{\arraystretch}{1.3}
	\centering
	\setlength\tabcolsep{5pt}
	\caption{Predicted longitudinal fraction ($R_L$) and $CP$-odd fraction ($R_\perp$) of state.}
	\label{tab15}
	\begin{tabular}{lcc}
		
		\hline
		&Longitudinal&\\
		&polarization&\ $CP$ -odd\\
		Decay modes &   fraction($R_L$)& \ fraction($R_\perp$)\\
			\hline ${B^-_c}\to J/\psi \rho^-$&0.932&0.018\\
		${B^-_c}\to \psi(2S) \rho^-$&0.891&0.024\\
		${B^-_c}\to \psi( 3S) \rho^-$&0.853&0.026\\
		\hline
		${B^-_c}\to J/\psi K^{*-}$&0.911&0.023\\
		${B^-_c}\to \psi(2S) K^{*-}$&0.860&0.030\\
		${B^-_c}\to \psi(3S) K^{*-}$&0.813&0.032\\
		\hline
		${B^-_c}\to J/\psi D_s^{*-}$&0.573&0.176\\
		${B^-_c}\to \psi(2S) D_s^{*-}$&0.405&0.353\\
		${B^-_c}\to \psi(3S) D_s^{*-}$&0.336&0.137\\
		\hline
		${B^-_c}\to J/\psi D^{*-}$&0.568&0.152\\
		${B^-_c}\to \psi(2S) D^{*-}$&0.379&0.246\\
		${B^-_c}\to \psi(3S) D^{*-}$&0.447&0.415\\
		\hline
		${B^-}\to J/\psi\rho^-$&0.757& 0.159\\
		${B^-}\to \psi(2S)\rho^-$&0.565&0.276\\
		${B^-}\to \psi(3S)\rho^-$&0.421& 0.345\\
		\hline
		${B^-}\to J/\psi K^{*-}$&0.728&0.141\\
		${B^-}\to \psi(2S) K^{*-}$&0.549&0.200\\
		${B^-}\to \psi(3S) K^{*-}$&0.426&0.201\\
		\hline
		${B^-}\to D^{*0}(1S)\rho^-$&0.945&0.013\\
		${B^-}\to D^{*0}(2S)\rho^-$&0.895&0.021\\
		${B^-}\to D^{*0}(3S)\rho^-$&0.849&0.024\\
		\hline
		${B^-}\to D^{*0}(1S)K^{*-}$&0.928&0.017\\
		${B^-}\to D^{*0}(2S)K^{*-}$&0.865&0.027\\
		${B^-}\to D^{*0}(3S)K^{*-}$&0.807&0.030\\
		\hline
		$\bar{B^0}\to D^{*+}(1S)\rho^-$&0.945&0.013\\
		$\bar{B^0}\to D^{*+}(2S)\rho^-$&0.899&0.020\\
		$\bar{B^0}\to D^{*+}(3S)\rho^-$&0.845&0.025\\
		\hline
		$\bar{B^0}\to D^{*+}(1S)K^{*-}$&0.928&0.017\\
		$\bar{B^0}\to D^{*+}(2S)K^{*-}$&0.870&0.026\\
		$\bar{B^0}\to D^{*+}(3S)K^{*-}$&0.802&0.031\\
		\hline
		$\bar{B^0_s}\to D_s^{+*}(1S)\rho^-$&0.945&0.013\\
		$\bar{B^0_s}\to D_s^{+*}(2S)\rho^-$&0.902&0.021\\
		$\bar{B^0_s}\to D_s^{+*}(3S)\rho^-$&0.852&0.025\\
		\hline
		$\bar{B^0_s}\to D_s^{+*}(1S) K^{*-}$&0.928&0.017\\
		$\bar{B^0_s}\to D_s^{+*}(2S) K^{*-}$&0.873&0.026\\
		$\bar{B^0_s}\to D_s^{+*}(3S) K^{*-}$&0.811&0.031\\ 
	\hline
	\end{tabular}
	\end{table}
\section{Summary and conclusion}
In this work, we study the exclusive two-body nonleptonic decays of $b$-flavored ($\bar{B^0}$, $\bar{B_s^0}$, $B^-$ and $B_c^-$) mesons to $S$-wave charmonium and charm meson $1^-$ states, in the framework of relativistic independent quark (RIQ) model. The weak decay form factors representing decay amplitude and their $q^2$-dependence are extracted from the overlapping integrals of the meson wave functions obtainable in the RIQ model. The predicted branching fractions for different decay modes are obtained in a wide range, from ${\cal O}(10^{-6})$ for $B_c^-$ decays to two charmful states to as high as $\sim 1.54 \%$, $\sim 1.42 \%$ and $\sim 1.17 \%$ for $B^-\to D^{*0}\rho^-$, $\bar{B^0}\to D^{*+}\rho^-$ and $\bar{B_s^0}\to D_s^{*+}\rho^-$, respectively. Our results are in general agreement with the available experimental data and other SM predictions. The decay modes with predicted branching fractions in the order: ${\cal O}(10^{-2})$, which include the $B_c$-meson decays to $1S$ charmonium and charm meson states as well as $\bar{B_s^0}$, $\bar{B^0}$, $B^-$ decays to two charmful mesons in their ground state, should be experimentally accessible. The decay modes to $2S$ and $3S$ charmonium states such as $B_c^-\to \psi(2S)\rho^-, \psi(2S)D_s^{*-}$, $\bar{B_s^0}\to D_s^{*+}(2S)\rho^-$, $\bar{B^0}\to D^{*+}(2S)\rho^-$, $B^-\to D^{*0}(2S)\rho^-$,  $B^-\to \psi(2S) K^{*-}$, $B^-\to \psi(2S) \rho^-$ and $\bar{B_s^0}\to D_s^{*+}(3S)\rho^-$, $\bar{B^0}\to D^{*+}(3S)\rho^-$, $B^-\to D^{*0}(3S)\rho^-$, $B^-\to \psi(3S) K^{*-}$, $B^-\to \psi(3S) \rho^-$ with predicted branching fractions upto $\sim10^{-4}$ may be accesible at high luminosity hadron colliders in near future. Other decay modes and especially $B_c$-decay to two charmful states with predicted branching factions in the order ${\cal O}(10^{-6})$ can not reach the detection ability of the current experiments. As expected, our predicted branching fractions are obtained in the hierarchy:
\begin{eqnarray}
  {\cal B}(B_F\to V_1(3S)V_2)&&<{\cal B}(B_F\to V_1(2S)V_2)\nonumber \\ &&<{\cal B}(B_F\to V_1(1S)V_2). \nonumber 
\end{eqnarray}
   
This is due to i) the nodal structure of the participating daughter mesons in their excited states, ii) tighter phase space and iii) typical $q^2$-dependence of the weak decay form factors for decay modes to higher excited ($2S$ and $3S$) states in comparison to that for the  corresponding decay modes to the corresponding ground $(1S)$ state.
\par The relative size of branching fractions is broadly estimated from a power counting of QCD factors: ($a_1, a_2$) in the Wolfenstein parametrization. The class I decay modes characterized by $a_1$ are found to have large branching fractions, as expected; compared to those obtained for class II decays which are determined by $a_2$. The branching fractions of class III decays characterized by Pauli interference for $B_c$-decays to two charmful states in particular, obtained in the order of magnitude ${\cal O}(10^{-6})$ can not be measured in current experiments.
\par In view of experimental data favoring a constructive interference of the color-favored and color-suppressed $b$-\\flavored meson decays, the effect of Pauli interference is studied in different decay modes, by evaluating the enhancement factor in each such decay mode. For ${B^-_c}\to \bar{D}^{*0}D_s^{*-}$, ${B^-_c}\to \bar{D}^{*0}(2S)D_s^{*-}$, ${B^-_c}\to \bar{D}^{*0}D^{*-}$ and ${B^-_c}\to \bar{D}^{*0}(2S)D^{*-}$, the enhancement (in $\%$) is found to be $\sim 63$, $\sim 96$, $\sim 44$ and $\sim 100$, respectively. This shows that the Pauli interference is more significant in $B_c^-$-decays to two charmful states: ${B^-_c}\to \bar{D}^{*0}(1S, 2S)D_{(s)}^{*-}$ compared to other decay modes. This is particularly important since such decay modes have been proposed for extracting the CKM angle $\gamma$ through amplitude relations. 
\par  We predict the longitudinal polarization fraction ($R_L$) and transverse polarization ($CP$-odd) fraction ($R_\perp$). We find that predicted $R_L$ dominates in all decay modes to two charmful states; considered in the present study. The $CP$-odd fraction ($R_\perp$) in nonleptonic $B_c^-$-decays to two charmful states are obtained one order magnitude higher than that in such other $b$-flavored meson decays; which indicates the $CP$- violation is more significant in $B_c^-\to D^{*0}D_{(s)}^{*-}$ compared to that obtained in ($\bar{B^0}$, $\bar{B_s^0}$ and $B^-$)-meson decays. For color-favored $B_c^-\to D^*(D^{*-}, D_s^{*-})$ decays, the effect arising due to the short-distance non-spectator contribution is shown to be marginal. However, the long-distance (LD) nonfactorizable contributions from rescattering effects, final-state interactions, etc., may not be negligible. If a significant LD effect exists, one expects a large $CP$-odd fraction in these decays. The predicted longitudinal and transverse helicity amplitudes and the form factor $g(q^2)$ yield $R_\perp$ values for different $B_F\to  V_1(nS)V_2$ decays. In particular, the predicted $R_\perp$ values for the transitions with two charmful final states indicate nonvanishing LD contributions, which lead to a significant CP-violation in $B_c^-\to D^{*0}D^{*-}_{(s)}$ and $\bar{D}^{*0}D^{*-}_{(s)}$ decays.
\par In conclusion, the present analysis shows that the factorization approximation works reasonably well in describing the exclusive nonleptonic $B_F\to V_1(nS)V_2$ decays in the framework of the RIQ model.

\appendix
\section{Quark orbitals and wave packet representation of the meson state in RIQ model framework}\label{app}
In the RIQ model, a meson is picturized as a color-singlet assembly of a quark and an antiquark independently confined by an effective and average flavor-independent potential in the form:
$U(r)=\frac{1}{2}(1+\gamma^0)(ar^2+V_0)$, where ($a$, $V_0$) are the potential parameters. It is believed that the zeroth-order quark dynamics  generated by the phenomenological confining potential $U(r)$ taken in equally mixed scalar-vector harmonic form can provide an adequate tree-level description of the decay process being analyzed in this work. With the interaction potential $U(r)$ put into the zeroth-order quark lagrangian density, the ensuing Dirac equation admits a static solution of positive and negative energy as: 
\begin{eqnarray}
	\psi^{(+)}_{\xi}(\vec r)\;&=&\;\left(
	\begin{array}{c}
		\frac{ig_{\xi}(r)}{r} \\
		\frac{{\vec \sigma}.{\hat r}f_{\xi}(r)}{r}
	\end{array}\;\right){{\chi}_{l,j,m_j}}(\hat r),
	\nonumber\\
	\psi^{(-)}_{\xi}(\vec r)\;&=&\;\left(
	\begin{array}{c}
		\frac{i({\vec \sigma}.{\hat r})f_{\xi}(r)}{r}\\
		\frac{g_{\xi}(r)}{r}
	\end{array}\;\right)\ {\tilde \chi}_{l,j,m_j}(\hat r),
\end{eqnarray}
where, $\xi=(nlj)$ represents a set of Dirac quantum numbers specifying 
the eigenmodes. $\chi_{ljm_j}(\hat r)$ and ${\tilde \chi}_{ljm_j}(\hat r)$
are the spin angular parts given by,
\begin{eqnarray}
	\chi_{ljm_j}(\hat r) &=&\sum_{m_l,m_s}<lm_l\;{1\over{2}}m_s|
	jm_j>Y_l^{m_l}(\hat r)\chi^{m_s}_{\frac{1}{2}},\nonumber\\
	{\tilde \chi}_{ljm_j}(\hat r)&=&(-1)^{j+m_j-l}\chi_{lj-m_j}(\hat r).
\end{eqnarray}
With the quark binding energy $E_q$ and quark mass $m_q$ written in the form $E_q^{\prime}=(E_q-V_0/2)$, $m_q^{\prime}=(m_q+V_0/2)$ and $\omega_q=E_q^{\prime}+m_q^{\prime}$, one can obtain solutions to the resulting radial equation for $g_{\xi}(r)$ and $f_{\xi}(r)$ in the form:
\begin{eqnarray}
	g_{nl}&=& N_{nl} \Big(\frac{r}{r_{nl}}\Big)^{l+1}\exp (-r^2/2r^2_{nl})
	L_{n-1}^{l+1/2}(r^2/r^2_{nl}),\nonumber\\
	f_{nl}&=&\frac{N_{nl}}{r_{nl}\omega_q} \Big(\frac{r}{r_{nl}}\Big)^{l}\exp (-r^2/2r^2_{nl})\nonumber\\
	&\times &\left[\Big(n+l-\frac{1}{2}\Big)L_{n-1}^{l-1/2}(r^2/r^2_{nl})
	+nL_n^{l-1/2}(r^2/r^2_{nl})\right ],
\end{eqnarray}
where, $r_{nl}= (a\omega_{q})^{-1/4}$ is a state independent length parameter, $N_{nl}$
is an overall normalization constant given by
\begin{equation}
	N^2_{nl}=\frac{4\Gamma(n)}{\Gamma(n+l+1/2)}\frac{(\omega_{nl}/r_{nl})}
	{(3E_q^{\prime}+m_q^{\prime})},
\end{equation}
and
$L_{n-1}^{l+1/2}(r^2/r_{nl}^2)$ etc. are associated Laguerre polynomials. The radial solutions yield an independent quark bound-state condition in the form of a cubic equation:
\begin{equation}
	\sqrt{(\omega_q/a)} (E_q^{\prime}-m_q^{\prime})=(4n+2l-1).
\end{equation}
From the solution of the cubic equation (A.5), the zeroth-order binding energies of 
the confined quark and antiquark are obtained for all possible eigenmodes. 
\par In the relativistic independent particle picture of this model, the relativistic constituent quark and antiquark are thought to move independently inside the meson bound-state $|B_F(\vec{p}, S_{B_F})\rangle$ with their momentum $\vec p_b$ and ${\vec{p}}_{\bar{q}}$, respectively. In order to study the decay process which takes place in the momentum eigenstates of participating mesons, we Fourier transform the quark orbitals (A.1) to momentum space and obtain the momentum probability amplitude of the quark and antiquark of participating mesons in the following forms:\\
\noindent For ground state mesons:($n=1$,$l=0$),
\begin{eqnarray}
	G_b(\vec p_b)&&={{i\pi {\cal N}_b}\over {2\alpha _b\omega _b}}
	\sqrt {{(E_{p_b}+m_b)}\over {E_{p_b}}}(E_{p_b}+E_b)\nonumber\\
	&&\times\exp {\Big(-{
			{\vec {p_b}}^2\over {4\alpha_b}}\Big)},\nonumber\\
	{\tilde G}_{\bar{q}}(\vec p_{\bar{q}})&&=-{{i\pi {\cal N}_{\bar{q}}}\over {2\alpha _{\bar{q}}\omega _{\bar{q}}}}
	\sqrt {{(E_{p_{\bar{q}}}+m_{\bar{q}})}\over {E_{p_{\bar{q}}}}}(E_{p_{\bar{q}}}+E_{\bar{q}})\nonumber\\
	&&\times\exp {\Big(-{
			{\vec {p}}_{\bar{q}}^2\over {4\alpha_{\bar{q}}}}\Big)}.
\end{eqnarray}
For the excited meson state:($n=2$,$l=0$),
\begin{eqnarray}
	G_b(\vec p_b)&&={{i\pi {\cal N}_b}\over {2\alpha _b\omega _b}}
	\sqrt {{(E_{p_b}+m_b)}\over {E_{p_b}}}(E_{p_b}+E_b)\nonumber\\
	&&\times \Big(\frac{\vec{p_b}^2}{2\alpha_b}-\frac{3}{2}\Big)\exp {\Big(-{
			{\vec {p_b}}^2\over {4\alpha_b}}\Big)},\nonumber\\
	{\tilde G}_{\bar{q}}(\vec p_{\bar{q}})&&=-{{i\pi {\cal N}_{\bar{q}}}\over {2\alpha _{\bar{q}}\omega _{\bar{q}}}}
	\sqrt {{(E_{p_{\bar{q}}}+m_{\bar{q}})}\over {E_{p_{\bar{q}}}}}(E_{p_{\bar{q}}}+E_{\bar{q}})\nonumber\\
	&&\times\Big(\frac{\vec{p}_{\bar{q}}^2}{2\alpha_{\bar{q}}}-\frac{3}{2}\Big)\exp {\Big(-{
			{\vec {p}}_{\bar{q}}^2\over {4\alpha_{\bar{q}}}}\Big)}.
\end{eqnarray}
For the excited meson state:($n=3$,$l=0$),
\begin{eqnarray}
	G_b(\vec p_b)&&={{i\pi {\cal N}_b}\over {2\alpha _b\omega _b}}
	\sqrt {{(E_{p_b}+m_b)}\over {E_{p_b}}}(E_{p_b}+E_b)\nonumber\\
	&&\times \Big(\frac{{\vec{p}_b}^4}{8\alpha_b^2}-\frac{5\vec{p_b}^2}{4\alpha_b}+\frac{15}{8}\Big)\exp {\Big(-{{\vec {p_b}}^2\over {4\alpha_b}}\Big)},\nonumber\\
	{\tilde G}_{\bar{q}}(\vec p_{\bar{q}})&&=-{{i\pi {\cal N}_{\bar{q}}}\over {2\alpha _{\bar{q}}\omega _{\bar{q}}}}
	\sqrt {{(E_{p_{\bar{q}}}+m_{\bar{q}})}\over {E_{p_{\bar{q}}}}}(E_{p_{\bar{q}}}+E_{\bar{q}})\nonumber\\
	&&\times \Big(\frac{{\vec{p}_{\bar{q}}}^4}{8\alpha_{\bar{q}}^2}-\frac{5\vec{p}_{\bar{q}}^2}{4\alpha_{\bar{q}}}+\frac{15}{8}\Big)\exp {(-{{\vec {p}}_{\bar{q}}^2\over {4\alpha_{\bar{q}}}})}.
\end{eqnarray}  
With the momentum probability amplitudes of quark constituents, we construct an effective momentum distribution function for the meson state in the form ${\cal G}_{B_F}(\vec{p}_b,\  \vec{p}_{\bar{q}})=\\\sqrt{G_b(\vec{p}_b) G_{\bar{q}}(\vec{p}_{\bar{q}})}$ in the light of an ansatz of Margolis and Mendel in their bag model description of the meson bound state \cite{A96}.
\par Here the effective momentum distribution function ${\cal G}_{B_F}\\(\vec{p}_b,\  \vec{p}_{\bar{q}})$, which in fact, embodies the bound-state characteristics of $|B_F(\vec{p},S_{B_F})\rangle$, defines the meson bound state at definite momentum $\vec{p}$ and spin projection $S_{B_F}$ in the form:
  \begin{eqnarray}
  |B_F(\vec{p},S_{B_F})\rangle&&={\hat{\Lambda}}(\vec{p},S_{B_F})|(\vec{p}_b,\lambda_b);(\vec{p}_{\bar{q}},\lambda_{\bar{q}})\rangle\nonumber\\&&={\hat{\Lambda}}(\vec{p},S_{B_F})\hat{b}^\dagger_b(\vec{p}_b,\lambda_b) \hat{\tilde{b}}^\dagger_{\bar{q}}(\vec{p}_{\bar{q}},\lambda_{\bar{q}})|0\rangle,
  \end{eqnarray}
where $|(\vec{p}_b,\lambda_b);(\vec{p}_{\bar{q}},\lambda_{\bar{q}})\rangle$ is the Fock-space representation of unbound quark and antiquark in a color-singlet configuration with respective momentum and spin: $(\vec{p}_b,\lambda_b)$ and $(\vec{p}_{\bar{q}},\lambda_{\bar{q}})$. $\hat{b}^\dagger_b(\vec{p}_b,\lambda_b)$ and $\hat{\tilde{b}}^\dagger_{\bar{q}}(\vec{p}_{\bar{q}},\lambda_{\bar{q}})$ are the quark and antiquark
creation operators and ${\hat{\Lambda}}(\vec{p},S_{B_F})$ is a bag-like integral operator taken in the form:
\begin{eqnarray}
 {\hat{\Lambda}}(\vec{p},S_{B_F})&&=\frac{\sqrt{3}}{\sqrt{N_{B_F}(\vec{p})}}\sum_{\lambda_b,\lambda_{\bar{q}}}\zeta_{b,\bar{q}}^{B_F}\int d^3\vec{p}_b\ d^3\vec{p}_{\bar{q}}\nonumber\\&& \delta^{(3)}(\vec{p}_b+\vec{p}_{\bar{q}}-\vec{p})\ {\cal G}_{B_F}(\vec{p}_b, \vec{p}_{\bar{q}}).   
\end{eqnarray}
Here, $\sqrt{3}$ is the effective color factor and $\zeta_{b,\bar{q}}^{B_F}$ is the $SU(6)$ spin-flavor coefficients for the $B_F$-meson state. Imposing the normalization condition in the form  
$\langle B_F(\vec{p})|B_F(\vec{p}^{'})\rangle=\\\delta^{(3)}{(\vec{p}-\vec{p}^{'})}$, the meson state normalization $N_{B_F}(\vec{p})$ is obtainable in an integral form
\begin{equation}
	N_{B_F}(\vec{p})=\int d^3\vec{p}_b \ |{\cal G}_{B_F}(\vec{p}_b,\  \vec p_{\bar{q}})|^2.
\end{equation}
\begin{acknowledgements}
The library and computational facilities provided by authorities of Siksha 'O' Anusandhan Deemed to be University, Bhubaneswar, 751030, India are duly acknowledged.
\end{acknowledgements}


\begin{thebibliography}{100}

\bibitem{A1}
	BELLE collaboration, Observation of $\eta_c(2S)$ in exclusive $B\to KK_SK^-\pi^+$ decays, {\it Phys. Rev. Lett.} {\bf 89} (2002) 102001 [hep-ex/0206002]. 
	\bibitem{A2} 
	BELLE collaboration, Observation of a new $D_{(sJ)}$ meson in $B^+\to \bar{D}^0D^0K^+$ decays, {\it Phys. Rev. Lett.} {\bf 100} (2008) 092001 [0707.3491].
	\bibitem{A3}
	LHCb collaboration, Determination of quantum numbers for several excited charmed mesons observed in $B^- \to D^{*+}\pi^-\pi^-$ decays, {\it Phys. Rev.} D {\bf 101} (2020) 032005 [1911.05957].
	\bibitem{A4}
	LHCb collaboration, Study of $D_J$ meson decays to $D^+\pi^-$, $D^0\pi^+$ and $D^{*+}\pi^-$ final states in $pp$ collision, {\it JHEP} {\bf 09} (2013) 145 [1307.4556].
	\bibitem{A5}
	 DELPHI collaboration, First evidence for a charm radially excitation, $D^{*'}$, {\it Phys. Lett.} B {\bf 426} (1998) 231.
    \bibitem{A6}
    LHCb collaboration, Amplitude analysis of $B^-\to D^+\pi^-\pi^-$ decays, {\it Phys. Rev.} D {\bf 94} (2016) 072001 [1608.01289].
    \bibitem{A7}
    T. Matsuki, T. Morii and K. Sudho, Radial Excitation of Heavy Mesons, {\it Eur. Phys. J.} A {\bf 31} (2007) 701 [hep-ph/0610186].
    \bibitem{A8}
    S. Godfrey and N. Isgur, Mesons in a Relativized Quark Model with Chromodynamics, {\it Phys. Rev.} D {\bf 32} (1985) 189.
    \bibitem{A9}
    BABAR collaboration, Branching fraction measurement of $\bar{B}^0\to D^{(*)+}\pi^-$ and $B^-\to D^{(*)0}\pi^-$ and isospin analysis of $\bar{B}\to D^{(*)}\pi$ decays, {\it Phys. Rev.} D {\bf 75} (2007) 031101 [hep-ex/0610027].
    \bibitem{A10}
    BELLE collaboration, Measurement of branching fraction ratios and $CP$ asymmetries in $B^{\pm}\to D_{CP}K^{\pm}$, {\it Phys. Rev.} D {\bf 68} (2003) 051101 [hep-ex/0304032].
    \bibitem{A11}
    LHCb collaboration, Measurement of branching fraction of decays $B_s^0\to D_s^{\mp}K^{\pm}$ and $B_s^0\to D_s^{-}\pi^+$, {\it JHEP} {\bf 06} (2012) 115 [1204.1237].
    \bibitem{A12}
    R. Aaij {\it et al.} [LHCb collaboration], {\it Phys. Rev.} D {\bf 87} (2013) no.11, 112012 Addendum: [{\it Phys. Rev.} D {\bf 89} (2014) no.1, 019901][arXiv:1304.4530[hep-ex]]. 
    \bibitem{A13}
    G. Aad {\it et al.} [ATLAS Collaboration],  {\it Eur. Phys. J.} C {\bf 76} (2016) no.1, 4 [arXiv:1507.07099[hep-ex]].
    \bibitem{A14}
    X. Liu, Z-J.Xiao and C-D.Lu, {\it Phys. Rev.} D {\bf 81}, 014022 (2010).
    \bibitem{A15}
    C.-H. Chang and Y.-Q. Chen, The decays of $B_{(c)}$ meson, {\it Phys. Rev.} D {\bf 49}  (1994) 3399.
    \bibitem{A16}
    C. Chang, H.-F. Fu, G.-L. Wang and J.-M. Zhang, Some of semileptonic and nonleptonic decays of $B_c$ meson in a Bethe-Salpeter relativistic quark model, {\it Sci. China Phys. Mech. Astron.} {\bf 58}, (2015) 071001 [1411.3428].
    \bibitem{A17}
    J.-F. Liu and K.-T. Chao, $B_c$ meson weak decays and $CP$-violation, {\it Phys. Rev.} D {\bf 56} (1997) 4133.
    \bibitem{A18}
    Tian Zhou {\it et al.} {\it J. Phys. G: Nucl. Part. Phys.} {\bf 48}, 055006 (2021).
    \bibitem{A19}
    Tian Zhou {\it et al.} {\it Eur. Phys. J.} C {\bf 81}, 339 (2021) .
    \bibitem{A20}
    D. Ebert, R. N. Faustov, and V. O. Galkin, {\it Phys. Rev.} D {\bf 68}, 094020 
    (2003).
    \bibitem{A21}
    M.A. Ivanov, J.G. Korner, P. Santorelli, {\it Phys. Rev.} D {\bf 71}, 094020 (2005).
    \bibitem{A22}
     M.A. Ivanov, J.G. Korner, P. Santorelli, {\it Phys. Rev.} D {\bf 63}, 074010 (2001).
    \bibitem{A23}
    M.A. Ivanov, J.G. Korner, P. Santorelli, {\it Phys. Rev.} D {\bf 73}, 054024 (2006).
    \bibitem{A24}
    Bediaga, I., and J. H. Munoz, Production of radially excited charmonium mesons in two-body nonleptonic $B_c$ decays; arXiv:1102.2190 (2011).
    \bibitem{A25}
    Hong-Wei Ke, Tan Liu, and Xue-Qian Li, {\it Phys. Rev.} D {\bf 89}, 017501 (2014).
    \bibitem{A26}
    R. N. Faustov and V. O. Galkin, {\it Phys. Rev.} D {\bf 87 }, 034033 (2013).
    \bibitem{A27}
    Pietro Colangelo and Fulvia De Fazio, {\it Phys. Rev.} D {\bf 61}, 034012 (2000).
    \bibitem{A28}
   M. Beneke, G. Buchalla, M. Neubert and C. T. Sachrajda, QCD factorization for exclusive, nonleptonic $B$ meson decays: General arguments and the case of heavy light final states, {\it Nucl. Phys.} B {\bf 591} (2000) 313 [hep-ph/0006124].
    \bibitem{A29}
    R.-H. Li, C.-D. Lu and H. Zou, The $B(B_{(s)})\to D_{(s)}P,  D_{(s)}V, D_{(s)}^*P$ and $D_{(s)}^*V$ decays in the perturbative QCD approach, {\it Phys. Rev.} D {\bf 78} (2008) 014018 [0803.1073].
    \bibitem{A30}
    G. Li, F.l. Shao and W. Wang,$B_s\to D_s(3040)$ form factors and $B_s$ decays into $D_s(3040)$, {\it Phys. Rev.} D {\bf 82} (2010) 094031 [1008.3696].
    \bibitem{A31}
      R.-H. Li, C.-D. Lu  and Y.-M. Wang, Exclusive $B_s$ decays to the charmed mesons $D_s^+(1968,2317)$ in the standard model, {\it Phys. Rev.} D {\bf 80} (2009) 014005 [0905.3259].
    \bibitem{A32}
    M. Neubert and B. Stech, Nonleptonic weak decays of $B$ mesons, hep-ph/9705292.
    \bibitem{A33}
    N, Barik, P. C. Dash, {\it Phys. Rev.} D {\bf 63}, 114002 (2001).
    \bibitem{A34}
    Sk. Naimuddin, S. Kar, M. Priyadarshini, N. Barik, P. C. Dash, {\it Phys. Rev.} D {\bf 86}, 094028 (2012).
    \bibitem{A35}
     N. Barik,1, Sk. Naimuddin,2, P. C. Dash and Susmita Kar, {\it Phys. Rev.} D {\bf 80} 014004 (2009). 
     \bibitem{A36}
     S. Kar, P. C. Dash, M. Priyadarsini, Sk Naimuddin, N. Barik, {\it Phys. Rev.} D {\bf 88}, 094014 (2013).
    \bibitem{A37}
    M. Wirbel, B. Stech, and M. Bauer, {\it Z. Phys.} C {\bf 29}, 637 (1985).
    \bibitem{A38}
    M. Bauer, B. Stech, and M. Wirbel, {\it Z. Phys.} C {\bf 34}, 103 (1987).
    \bibitem{A39}
    L.-L. Chau, H.-Y. Cheng, W. K. Sze, H. Yao,and B. Tseng, {\it Phys. Rev.} D {\bf 43}, 2176 (1991).
    \bibitem{A40}
    L.-L. Chau, H.-Y. Cheng, W. K. Sze, H. Yao,and B. Tseng, {\bf 58}, 019902 (1998).
    \bibitem{A41}
    I. P. Gouz, V. V. Kiselev, A. K. Likhoded, V. I. Romanovsky,
    and O. P. Yushchenko, {\it Phys. At. Nucl.} {\bf 67}, 1559 (2004).
    \bibitem{A42}
    M. Beneka, G. Buchalla, M. Neubert, and C. T. sachrajda, {\it Phys. Rev. Lett.} {\bf 83} 1914 (1999).
    \bibitem{A43}
     M. Beneka, G. Buchalla, M. Neubert, and C. T. sachrajda, {\it Nucl. Phys.} B {\bf 591} 313 (2000).
     \bibitem{A44}
     M. Beneka, G. Buchalla, M. Neubert, and C. T. sachrajda, {\it Nucl. Phys.} B {\bf 606}, 245 (2001).
    \bibitem{A45}
    M. Beneka and M. Neubert, {\it Nucl. Phys.} B {\bf 675}, 33 (2003).
    \bibitem{A46}
    C. E. Thomas, {\it Phys. Rev.} D {\bf 73}, 054016 (2006).
    \bibitem{A47}
    V. V. Kiselev, A. E. Kovalsky, and A. K. Likhoded, {Phys. At. Nucl.} {\bf 64}, 1860 (2001).
    \bibitem{A48}
    V. V. Kiselev, A. E. Kovalsky, and A. K. Likhoded, {\it Nucl. Phys.} B {\bf 585}, 353 (2000).
     \bibitem{A49}
    V. V. Kiselev:arXiv: hep-ph/0211021 (2002).
    \bibitem{A50}
     X. Q. Yu and X. L. Zhou, {\it Phys. Rev.} D {\bf 81}, 037501 (2010).
    \bibitem{A51}
     J. D. Bjorken, in Proceedings of the International Workshop, Crete, Greece, 1988, edited by G. Branco and J. Reeo, Development in High Energy Physics {\it Nucl. Phys.} B Proc. Suppl. {\bf 11 }, 325 (1989).
    \bibitem{A52}
    A. J. Buras, J. M. Gerard, and R. Ruckl, {\it Nucl. Phys.} B {\bf 268}, 16 (1986).
    \bibitem{A53}
    M. Neubert, {\it Phys. Rep.} {\bf 245}, 259 (1994).
    \bibitem{A54}
    G. Buchalla et al., {\it Eur. Phys. J.} C {\bf 57}, 309 (2008); A. J. Buras, arXiv:hep-ph/9806471.
    \bibitem{A55}
    Lopamudra Nayak , P. C. Dash , Susmita Kar and N. Barik, {\it Phys. Rev.} D {\bf 105}, 053007 (2022).
    \bibitem{A56}
    R. Aajj et al. (LHCb Collaboration), {\it J. High Energy Phys.} {\bf 09 }
    (2013) 075.
     \bibitem{A57}
     N. Brambilla {\it et al.} (Quarkonium Working Group), arXiv:hep-ph/0412158.
    \bibitem{A58}
   M. Neubert and B. Stech, {\it Adv. Ser. Dir. High Energy Phys.} {\bf 17}, 294 (1998).
    \bibitem{A59}
    H. M. Choi and C. R. Ji, {\it Phys. Rev.} D  {\bf 80}, 114003 (2009).
    \bibitem{A60}
    J. Sun, D. Du and Y. Yang, {\it Eur. Phys. J.} C {\bf 60}, 107 (2009).
    \bibitem{A61}
    J. Sun, Y. Yang, W. Du, and H. Ma, {\it Phys. Rev.} D {\bf 77} , 114004 (2008).
    \bibitem{A62}
    J. Sun, G. Xue, Y. Yang, G. Lu and D. Du, {\it Phys. Rev.} D {\bf 77}, 074013 (2008).
\bibitem{A63}
P. Colangelo and F. De Fazio, {\it Phys. Rev.} D {\bf 61}, 034012
(2000).
\bibitem{A64}
N. Sharma and R. C. Verma, {\it Phys. Rev.} D  {\bf 82}, 094014 (2010).
\bibitem{A65}
N. Sharma, R. Dhir, and R. C. Verma, {\it J. Phys.} G {\bf 37}, 075013 (2010).
\bibitem{A66}
R. Dhir and R. C. Verm, {\it Phys. Rev.} D  {\bf 79}, 034004 (2009).
\bibitem{A67}
N. Sharma, {\it Phys. Rev.} D  {\bf 81}, 014027 (2010).
\bibitem{A68}
   N. Barik, Sk. Naimuddin, P. C. Dash and Susmita Kar, {\it Phys. Rev.} D  {\bf 77}, 014038 (2008). 
   \bibitem{A69}
   N. Barik, Sk. Naimuddin, P. C. Dash and Susmita Kar, {\it Phys. Rev.} D  {\bf 78}, 114030 (2008).
   \bibitem{A70}
   N. Barik, Sk. Naimuddin and P. C. Dash, {\it Int. J. Mod. Phys.} A {\bf 24}, 2335 (2009).
   \bibitem{A71} 
    P. Colangelo, F. De Fazio and G. Nardulli, {\it Phys. Lett.} B {\bf 386}, 328 (1996).
    \bibitem{A72}
    G. Altarelli, N. Cabibbo, G. Corbo, L. Maiani and G. Martinelli, {\it Nucl. Phys.} B {\bf 208}, 365 (1982).
    \bibitem{A73}
    N. Barik and B. K. Dash, {\it  Phys. Rev.} D {\bf 33}, 1925 (1986).
    \bibitem{A74}
    N. Barik, B. K. Dash, and P. C. Dash,{\it  Pramana J. Phys.} {\bf 29}, 543 (1987).-
    \bibitem{A75}
    N. Barik and P. C. Dash, {\it Phys. Rev.} D {\bf 47}, 2788 (1993).
    \bibitem{A76}
     N. Barik, P. C. Dash, and A. R. Panda, {\it Phys. Rev.} D {\bf 46}, 3856 (1992).
     \bibitem{A77}
     N. Barik and P. C. Dash, {\it Phys. Rev.} D {\bf 49}, 299 (1994).
     \bibitem{A78}
     M. Priyadarsini, P. C. Dash, S. Kar, S. P. Patra and N. Barik, {\it Phys. Rev.} D {\bf 94}, 113011 (2016).
     \bibitem{A79}
     N. Barik and P. C. Dash, {\it Mod. Phys. Lett.} A {\bf 10}, 103 (1995).
     \bibitem{A80}
     N. Barik, S. Kar, and P. C. Dash, {\it Phys. Rev.} D {\bf 57}, 405 (1998).
     \bibitem{A81}
     N. Barik, Sk. Naimuddin, S. Kar, and P. C. Dash, {\it Phys. Rev.} D {\bf 63}, 014024 (2000).
   \bibitem{A82}
   R.L. Workman {\it et. al.}, (Particle Data Group), {\it Prog. Theor. Expt. Phys.} {\bf 2022}, 283C01 (2022).
   \bibitem{A83}
   S. Godfrey and K. Moats, {\it Phys. Rev.} D {\bf 93}, 034035 (2016).
   \bibitem{A84}
   Chen. Ying, Andrei Alexandru, Terrence Draper, Keh-Fei Liu, Zhaofeng Liu, and Yi-Bo Yang, Leptonic Decay Constant of $\rho$ at Physical Point [arXiv:1507.02541[hep-ex]] (2015).
   \bibitem{A85}
   Patricia Ball and Roman Zwicky, {\it Phys. Rev.} D {\bf 71}, 014029 (2005).
   \bibitem{A86}
   Guo-Li Wang, {\it Phys. Lett.} B {\bf 633}, 492 (2006). 
   \bibitem{A87}
   T. E. Browder and K. Honscheid, {\it Prog. Part. Nucl. Phys. } {\bf 35},
   81 (1995).
   \bibitem{A88}
   M. Neubert, V. Rieckert, B. Stech, and Q. P. Xu,
   in Heavy Flavors, edited by A. J. Buras and H. Lindner
   (World Scientific, Singapore, 1992) and references therein. 
   \bibitem{A89} 
    S. Dubnika, A. Z. Dubnikov, A. Issadykov, M. A. Ivanov and A. Liptaj, {\it Phys. Rev.} D {\bf 96}, 076017 (2017).   
    \bibitem{A90}
     L. Wolfenstein, {\it  Phys. Rev. Lett.} {\bf 51}, 1945 (1983).
    \bibitem{A91}
    V. V. Kiselev, {\it J. Phys.} G {\bf 30}, 1445 (2004).
    \bibitem{A92}
    R. Fleischer and D. Wyler, {\it Phys. Rev.} D  {\bf 62}, 057503 (2000).
    \bibitem{A93}
    M. Masetti,
    {\it Phys. Lett.} B {\bf 286}, 160 (1992).
    \bibitem{A94}
    M. A. Ivanov, J. G. Korner,
    and O. N. Pakhomova, {\it Phys. Lett.} B {\bf 555}, 189 (2003). 
    \bibitem{A95}
    H. N. Li and S. Mishima, {\it Phys. Rev.} D {\bf 71}, 054025 (2005).
 \bibitem{A96}
   B. Margolis and R. R. Mendel, {\it Phys. Rev.} D {\bf 28}, 468 (1983).
\end{thebibliography}
\end{document}